\documentclass[aps, prd, twocolumn, amsmath, amssymb, floats, floatfix, groupedaddress, nofootinbib, nopacs,notitlepage]{revtex4-1}

\usepackage{latexsym}
\usepackage{graphicx}
\usepackage{amsmath}
\usepackage{amssymb}
\usepackage{amsfonts}
\usepackage{times}
\usepackage{xspace} 
\usepackage[usenames]{color}
\usepackage{dcolumn}
\usepackage{bm}
\usepackage{mathrsfs}
\usepackage{appendix}
\usepackage{breqn}
\usepackage[]{units}       

\usepackage{subfigure}		
\usepackage[normalem]{ulem} 

\usepackage[tracking=true]{microtype}
\SetTracking{}{500}
\SetTracking{encoding={*}, shape=sc}{40}

\makeatletter
\let\cat@comma@active\@empty
\makeatother

\begin{document} 
 
\title{Hair-dressing Horndeski: an approach to hairy solutions in cubic Horndeski gravity}

\author{Reginald Christian Bernardo}
\email{rbernardo@nip.upd.edu.ph}
\author{Ian Vega}
\email{ivega@nip.upd.edu.ph}
\affiliation{National Institute of Physics, University of the Philippines, Diliman, Quezon City 1101, Philippines}
\date{\today}

\begin{abstract}
In obtaining exact solutions in gravitational theories containing arbitrary model functions, such as Horndeski gravity, one usually starts by prescribing the model functions of the theory and then goes on to solving their corresponding field equations. In this paper, we explore the extent to which the reverse procedure can be useful, whereby one starts with desired solutions and then determines the models that support them. Working within the phenomenologically interesting cubic and shift-symmetric sector of Horndeski gravity, we develop a method for obtaining exact static and spherically-symmetric solutions, one of which happens to be a new hairy black hole. We study this black hole and its properties. We also discuss the limitations of the method and its  possible extension to other Horndeski sectors.
\end{abstract}

\maketitle

\section{Introduction}
\label{sec:intro}

The detection of gravitational waves by LIGO is a spectacular addition to the illustrious list of observational successes supporting Einstein's general relativity (GR) as the correct theory of gravity on a wide range of scales \cite{bh_physics_community, gw_170817_ligo, gw_1_ligo,gw_2_ligo,gw_3_ligo,gw_4_ligo}. On the other hand, perplexing mysteries such as 
the accelerating expansion of the universe and the protracted quest for a direct detection of dark matter, notwithstanding the well-known incompatibility of quantum mechanics with general relativity at extremely small scales and high energies, all continue to inspire efforts to construct alternative theories of gravity \cite{accelerating_universe_reiss,accelerating_universe_tonry, dark_energy_frieman, dark_energy_weinberg,weinberg_cc,qg_deser1, qg_deser2, qg_hawking}. A happy compromise rests in the belief that general relativity is correct as an effective theory. But even so, the extrapolation of GR's validity to all scales can strike even the most credulous as rather unsatisfactory. Arguably the simplest modification to general relativity is a scalar-tensor theory where a gravitational degree of freedom is attached to at least one scalar field in addition to those that are already associated with the metric \cite{alternative_gravity_clifton, alternative_gravity_joyce, alternative_gravity_koyama}. The most general scalar-tensor theory with second-order field equations and a single scalar field was constructed by Horndeski way back in 1974, but has only faily recently attracted a lot of interest \cite{st_horndeski_seminal,st_horndeski_revive_deffayet, st_honrdeski_revive_deffayet_2, st_horndeski_galileons_charmousis, st_horndeski_cosmology_kobayashi}. Among its attractive features is that it encompasses GR, previous scalar-tensor theories such as Brans-Dicke and Einstein-Gauss-Bonnet gravity, and it can account for cosmic acceleration \cite{st_dark_energy_tsujikawa, horndeski_review_kobayashi2019}. The theory also remains phenomenologically viable, in spite of severe constraints set by LIGO's detection of gravitational waves \cite{dark_energy_derham, dark_energy_creminelli, dark_energy_ezquiaga, st_horndeski_vainshtein_dima, st_horndeski_cosmology_baker, st_horndeski_cosmology_sakstein, st_horndeski_cosmology_bettoni,st_horndeski_lss_peirone,
lombriser2016breaking, lombriser2017challenges}.

The strong-gravity regime of any alternative theory of gravity, particularly its black holes, is inherently interesting because it is where the strongest deviations from GR are expected. This regime is therefore where one expects the best chance to observationally distinguish GR from its competitors. Ironically, black holes in many alternative theories have a tendency to look very much like their GR counterparts. For shift-symmetric scalar-tensor theory in particular, a number of no-hair theorems have been discovered that explicitly rule out the possibility of obtaining black holes that support nontrivial scalar fields (or ``hair'') \cite{st_no_hair_theorem_hui, st_no_hair_benkel, st_no_hair_theorem_sotiriou_1, st_no_hair_theorem_sotiriou_2}. These no-hair theorems are of course limited by the assumptions that enter their proofs, and so outside these assumptions, it is certainly possible to obtain hairy black holes \cite{st_horndeski_babichev, st_horndeski_solutions_babichev_2, st_horndeski_cosmological_tuning_babichev, st_horndeski_slow_rotation_bh_maselli, st_horndeski_neutron_stars_maselli, st_horndeski_solutions_kobayashi, st_horndeski_solutions_babichev_0, st_horndeski_solutions_rinaldi, st_horndeski_solutions_anabalon, st_horndeski_solutions_minamitsuji, st_horndeski_solutions_gaete, achour2018stealth, antoniou2018evasion, antoniou2018black}. The complexity of the field equations generally prevent access to explicit analytical solutions, even static and spherically-symmetric ones. And when exact solutions can be found, they are restricted to particular sectors of theory \cite{st_horndeski_babichev, st_horndeski_solutions_babichev_2, st_horndeski_cosmological_tuning_babichev, st_horndeski_slow_rotation_bh_maselli, st_horndeski_neutron_stars_maselli, st_horndeski_solutions_kobayashi, st_horndeski_solutions_babichev_0, st_horndeski_solutions_rinaldi, st_horndeski_solutions_anabalon, st_horndeski_solutions_minamitsuji, st_horndeski_solutions_gaete, achour2018stealth, antoniou2018evasion, antoniou2018black}. Exact solutions to a large swath of Horndeski theory thus remain unavailable. Given the essential role exact solutions play in theory and practice as templates on which to build realistic (e.g. perturbative) solutions, the search for them continues to remain an active area of investigation.

A natural first step for constructing exact solutions is to realize that the fields and sources appear in different places in the field equations. So in principle, one can then just substitute trial fields into the field equations and then simply determine the sources needed to support them. Of course the problem with this approach is that it often gives unphysical solutions. In GR, one may simply plug-in any metric into the Einstein equation and build an appropriate stress-energy tensor to match the substituted metric; however, the stress-energy tensor built in this way will, in general, violate cherished energy conditions or possess singularities. Similarly, in electromagnetism, one may use Maxwell's equations to assign charges and currents to any electric and magnetic fields but the resulting solution then typically describes an unphysical distribution of charges (e.g. one not satisfying the continuity equation). In vacuum, where there are no sources to which one can attribute the fields, the strategy is much more limited. But certain alternative gravity theories like Horndeski theory are characterized by a number of free model functions, which can stand in as surrogates to matter fields. However, these model functions enter the field equations in a highly nontrivial way. As far as we know, the extent to which one can assign a set of model functions to generate a specified solution (i.e. gravitational field) has not yet been explicitly worked out. We are thus led to the question that drives this work: to what extent can we play this game in Horndeski theory? \footnote{The idea is similar in spirit to what is already well-established in quintessence cosmology wherein it is always possible to assign a scalar field potential to support any cosmological evolution.} Some progress on this is presented in Ref. \cite{st_horndeski_solutions_minamitsuji_2} to build stealth Schwarzschild black hole solutions in quadratic and cubic Horndeski theories and in Ref. \cite{st_horndeski_cosmology_emond2018} to build cosmological solutions.

For concreteness, we focus on static and spherically-symmetric solutions in the shift-symmetric sector of Horndeski gravity. Shift-symmetric scalars are also known as galileons. In this sector, a few hairy solutions have been obtained by a direct attack on the field equations  \cite{st_horndeski_babichev, st_horndeski_solutions_babichev_2, st_horndeski_cosmological_tuning_babichev, st_horndeski_slow_rotation_bh_maselli, st_horndeski_neutron_stars_maselli, st_horndeski_solutions_kobayashi, st_horndeski_solutions_babichev_0, st_horndeski_solutions_rinaldi, st_horndeski_solutions_anabalon, st_horndeski_solutions_minamitsuji, st_horndeski_solutions_gaete}, all the while exploting loopholes in the no-hair theorem for galileons \cite{st_no_hair_theorem_hui, st_no_hair_benkel, st_no_hair_theorem_sotiriou_1, st_no_hair_theorem_sotiriou_2}.

In this paper, we offer a different method for obtaining solutions in the shift-symmetric sector of Horndeski. Our approach provides a relatively easy path towards exact solutions, one of which will be shown to be a black hole. This black hole solution is quite different from the Schwarzschild geometry. To the best of our knowledge, there does not exist any other non-GR black hole solution for this Horndeski sector in the extant literature\footnote{Very recently, as just as this paper was being prepared, hairy stealth solutions (i.e. Schwarzschild metric) in cubic Horndeski were reported \cite{st_horndeski_solutions_minamitsuji_2}.}. The solution we present here is not a hairy Schwarzschild black hole. 

We specifically restrict our attention to the cubic sector for a number of reasons, the technical one being that its model function appears in only one place in the field equations of this sector. Apart from this, the cubic sector is quite appealing since its static and spherically symmetric solutions are known to be free from ghost and gradient instabilities under perturbations of the metric and the scalar field \cite{perturbation_kobayashi_1, perturbation_kobayashi_2}. It is worth mentioning that the cubic sector escapes the latest observational bounds on the speed of gravitational waves \cite{gw_170817_ligo, dark_energy_creminelli, dark_energy_ezquiaga}, though some works also claim that its cubic galileon limit (i.e, for which $G_3 \sim X$) is inconsistent with the cosmic microwave background, baryon acoustic oscillations, and weak lensing \cite{st_horndeski_galileon_barreira_1, st_horndeski_galileon_barreira_2,st_horndeski_renk, st_horndeski_galileon_peirone}. Recent work has also shown that Horndeski theory remains phenomenologically viable even when cosmological observations are taken into consideration \cite{st_horndeski_constraint_mancini2019, horndeski_constraint_noller2018, st_horndeski_constraint_komatsu2019}.

The rest of the paper proceeds as follows. To set the stage, we first provide a brief overview of Horndeski theory. In Section \ref{sec:no_hair}, we derive the necessary conditions for some version of the no-hair theorem, which states that no GR-like black holes can exist with the vanishing scalar current constraint in the sector. These necessary conditions are then exploited in Section \ref{sec:no_hair} to associate exact solutions with the cubic model function. A key insight here is that the model function can depend only on the kinetic density. In Sections \ref{sec:solutions_cubic_horndeski}, \ref{sec:recipe_regular}, and \ref{sec:black_hole}, we showcase our recipe for building exact analytical solutions, including a black hole solution, and their cubic Horndeski theories. We also discuss the asymptotic properties of the black hole. In Section \ref{sec:discussion}, we comment on the uniqueness of the model assigned to the hairy spacetime as obtained from the method. Finally, we conclude the paper with a discussion of the limitations of the method and its potential applicability to the wider sector of Horndeski theory. 

We work with a mostly plus signature $(-,+,+,+)$ and adopt geometric units in which $G = c = 1$.

\section{Cubic Horndeski theory}
\label{sec:horndeski_theory}

Horndeski theory or generalized galileon theory \cite{alternative_gravity_clifton, alternative_gravity_joyce, alternative_gravity_koyama, st_horndeski_seminal} is described by the action
\begin{equation}
\label{eq:horndeski_theory}
S = \int d^4 x \sqrt{-g} \left[ L_2 + L_3 + L_4 + L_5 \right]
\end{equation}
where
\begin{eqnarray}
L_2 &=& G_2 \\
L_3 &=& -G_3 \Box \phi \\
L_4 &=& G_4  R + G_{4X} \left[ \left( \Box \phi  \right)^2 - \left( \nabla_\mu \nabla_\nu \phi \right)^2  \right] \\
L_5 &=& G_5 G_{\mu\nu} \left( \nabla^\mu \nabla^\nu \phi \right) \\
& & - \frac{1}{6} G_{5X} \bigg[ \left( \Box \phi  \right)^3- 3 \Box \phi \left( \nabla_\mu \nabla_\nu \phi \right)^2 + 2 \left( \nabla_\mu \nabla_\nu \phi \right)^3  \bigg] \nonumber
\end{eqnarray}
where $G_i = G_i\left( \phi, X  \right)$ are arbitrary functions of the scalar field $\phi$ and its kinetic density $X = - g^{\alpha \beta} \left(\partial_\alpha \phi \right) \left( \partial_\beta \phi \right) / 2$. The shift-symmetric sector of the theory corresponds to $G_i \left( \phi, X \right) = G_i \left( X \right)$ and in this case the vacuum equations of motion of the metric and scalar fields can be written as
\begin{eqnarray}
\label{eq:covariant_einstein_equation} G_{\alpha \beta} - 8 \pi T_{\alpha \beta}^{(\phi)} &=& 0 \\
\label{eq:covariant_scalar_equation} \nabla_\alpha J^\alpha (x) = 0
\end{eqnarray} 
where $T_{\alpha \beta}^{(\phi)}$ is the stress-energy tensor of the scalar and $J^\alpha$ is the Noether current arising from the shift-symmetry $\phi \rightarrow \phi + \phi_0$ where $\phi_0$ is a constant. Being the most general scalar-tensor theory that evades Ostrogradski instabilities \cite{alternative_gravity_clifton, alternative_gravity_joyce, alternative_gravity_koyama}, Horndeski theory offers a very rich phenomenology. For instance, its covariant galileon limit is particularly interesting because it can account for cosmic acceleration without the need for a cosmological constant \cite{st_galileon_inflation_kobayashi, st_galileon_inflation_kobayashi_2, st_galileon_inflation_burrage, st_galileon_inflation_deffayet}. 

Current constraints on the theory include LIGO's observation that gravitational waves travel at the speed of light with an uncertainty of one part in $10^{15}$ \cite{gw_170817_ligo}. This strongly disfavors most of the quartic and quintic sectors (terms with $G_4$ and $G_5$) of the theory while leaving the quadratic and cubic sectors (terms with $G_2$ and $G_3$), all of which predict tensor perturbations that propagate at the speed of light, such as GR, quintessence and $k$-essence, kinetic gravity braiding, and Brans-Dicke sectors, unconstrained \cite{dark_energy_derham, dark_energy_creminelli, dark_energy_ezquiaga, st_horndeski_vainshtein_dima, st_horndeski_cosmology_baker, st_horndeski_cosmology_sakstein, st_horndeski_cosmology_bettoni}. Previous works have shown that the cubic galileon sector ($G_3 \sim X$) is inconsistent with the cosmic microwave background, baryon acoustic oscillations, and the integrated-Sachs-Wolfe effect. It has also been shown that the covariant galileon model ($G_2 \sim G_3 \sim X$ and $G_4 \sim G_5 \sim X^2$) is statistically ruled out by cosmological data including weak lensing \cite{st_horndeski_galileon_barreira_1, st_horndeski_galileon_barreira_2,st_horndeski_renk, st_horndeski_galileon_peirone}.  The rest of the quadratic and cubic sectors remain phenomenologically interesting when cosmological data is taken into consideration \cite{st_horndeski_constraint_mancini2019, horndeski_constraint_noller2018, st_horndeski_constraint_komatsu2019}. The current constraints on the Horndeski theory has definitely shifted some attention to the quadratic and cubic sectors of the theory \cite{st_horndeski_cubic_appleby, st_horndeski_scaling_solutions_albuquerque, st_horndeski_solutions_minamitsuji_2, st_horndeski_qnm_tattersall, st_dark_energy_tsujikawa,no_run_linder}.

For the rest of the paper, we shall focus on the Horndeski sector defined by the action
\begin{equation}
\label{eq:theory}
S = \int d^4 x \sqrt{-g} \left[ R + \left( X + L \right) - G\left(X\right) \Box \phi  \right]
\end{equation}
or, equivalently, by
\begin{eqnarray}
\label{eq:power_G2} G_2 &=& X + L \\
\label{eq:power_G3} G_3 &=& G(X) \\
\label{eq:power_G4} G_4 &=& 1
\end{eqnarray}
where $L$ is a cosmological constant and $G$ is an arbitrary function of the kinetic density that we'll refer to as the cubic model function. The stress-energy tensor and the scalar current given by Eqs. (\ref{eq:stress_energy}) and (\ref{eq:current_covariant}) can be derived using this action. In addition to this theory being phenomenologically interesting, it is as analytically tractable as single-scalar field cosmology since there is only a single free function entering the field equations. The simplicity that comes with this allows us to build a recipe for constructing static and spherically-symmetric solutions and a new hairy black hole solution.

\section{Staticity and spherical symmetry in cubic Horndeski}
\label{sec:no_hair}

In this section, we develop the field equations of Horndeski gravity to obtain convenient necessary conditions that any static and spherically-symmetric solution must satisfy. We then use them to prove an extension to the no-hair theorem for black holes in cubic Horndeski. The no-hair theorem is evaded by requiring that the scalar current, $J^\alpha$, vanishes. We show here that even within this space of vanishing-current solutions, an important class of solutions cannot support any scalar hair. This class includes, in particular, stealth (i.e. GR-like) black hole solutions such as the Schwarzschild geometry.

\subsection{Necessary conditions for solutions}

We consider static and spherically-symmetric scalar and tensor fields
\begin{eqnarray}
\label{eq:hair_static_spherical} \phi &=& \phi(r) \\
\label{eq:metric_static_spherical} ds^2 &=& - h(r) dt^2 + \frac{dr^2}{f(r)} + r^2 d\Omega^2
\end{eqnarray}
where $d\Omega^2$ is the metric on the unit two-sphere. The task is to solve the vacuum field equations of the cubic shift-symmetric Horndeski gravity where the scalar field's stress-energy tensor, $T^{(\phi)}_{\alpha \beta}$, and the scalar current, $J^\alpha$, are given by Eqs. (\ref{eq:stress_energy}) and (\ref{eq:current_covariant}), respectively. The $tt$- and $rr$-components of the Einstein equation are all that we need to deal with since the Bianchi identity, $\nabla_\beta G^{\alpha \beta} = 0$, and spherical symmetry guarantee that we can obtain $G_{\theta \theta} = G_{\phi \phi} / \sin^2\theta$ in terms of $G_{tt}$ and $G_{rr}$. The only nonvanishing component of the scalar current (Eq. (\ref{eq:current_covariant})) is the radial component which is
\begin{equation}
\label{eq:current}
J^r = \phi^\prime \left[ -f + \frac{1}{2} f^2 \phi^{\prime} \left( \frac{h^\prime}{h} + \frac{4}{r} \right) G_{X} \right] 
\end{equation}
where $G_X = d G/ dX$ and a prime denotes differentiation with respect to $r$. The no-hair theorem for the galileon derives from the argument that the radial component of the current vanishes for shift-symmetric Horndeski \cite{st_no_hair_theorem_hui}. The root $\phi' = 0$ of Eq. (\ref{eq:current}) is chosen for this no-hair branch. A loophole of the no-hair comes from exploiting the other roots $\phi' \neq 0$, i.e. the zeros of the expression in the square bracket in Eq. (\ref{eq:current}), which occurs when $\phi' G_X$ depends inversely on $\phi'$ \cite{st_no_hair_theorem_sotiriou_1, st_no_hair_theorem_sotiriou_2}.
 The no-hair theorem for the galileon can be evaded by setting the radial component of the scalar current to zero for all radii for models which put scalar field derivatives in the denominator \cite{st_horndeski_babichev, st_horndeski_solutions_babichev_2, st_horndeski_cosmological_tuning_babichev, st_horndeski_slow_rotation_bh_maselli, st_horndeski_neutron_stars_maselli, st_horndeski_solutions_kobayashi, st_horndeski_solutions_babichev_0, st_horndeski_solutions_rinaldi, st_horndeski_solutions_anabalon, st_horndeski_solutions_minamitsuji, st_horndeski_solutions_gaete, st_no_hair_theorem_hui, st_no_hair_benkel, st_no_hair_theorem_sotiriou_1, st_no_hair_theorem_sotiriou_2}. Setting the radial component of the the current to zero renders the scalar field equation, Eq.~(\ref{eq:covariant_scalar_equation}), trivially satisfied and forces the scalar field (or equivalently the kinetic density $X$) to assume a specific form in terms of the metric functions. Solutions of vanishing scalar current have been the focus of a number of attempts at building hairy black hole solutions \cite{st_horndeski_babichev, st_horndeski_solutions_babichev_2, st_horndeski_cosmological_tuning_babichev, st_horndeski_slow_rotation_bh_maselli, st_horndeski_neutron_stars_maselli, st_horndeski_solutions_kobayashi, st_horndeski_solutions_babichev_0, st_horndeski_solutions_rinaldi, st_horndeski_solutions_anabalon, st_horndeski_solutions_minamitsuji, st_horndeski_solutions_gaete}, and here we follow suit. From Eq. (\ref{eq:current}), we can then write down
\begin{equation}
\label{eq:current_constraint_1}
G_X \sqrt{-X} \sqrt{\frac{f}{2}} \left( \frac{h^\prime}{h} + \frac{4}{r} \right) = 1.
\end{equation}
At this point we make the important realization -- which is central to our approach -- that Eq. (\ref{eq:current_constraint_1}) requires that the particular combination of metric functions 
\begin{equation}
\sqrt{\frac{f}{2}} \left( \frac{h^\prime}{h} + \frac{4}{r} \right) 
\end{equation}
must be expressible entirely as a function of the kinetic density $X$. That is, there must exist a function $F(X)$ such that
\begin{equation}
F(X) = \sqrt{\frac{f}{2}} \left( \frac{h^\prime}{h} + \frac{4}{r} \right),
\end{equation}
or
\begin{equation}
\label{eq:prayer_function}
X = F^{-1} \left( \sqrt{\frac{f}{2}} \left( \frac{h^\prime}{h} + \frac{4}{r} \right) \right)
\end{equation} 
where $F^{-1}$ is the inverse of $F$. Of course, this can only happen if we can find an expression for the kinetic density in terms of the metric functions only. This turns out to be true quite often, but occassionally, such as the case of stealth (i.e. GR-like) black holes, $F^{-1}$ is a constant , and the inverse does not exist. In the next subsection, we set out to show this explicitly. The constant $F^{-1}$ case is the only case that completely hinders the possibility of assigning an inverse, thus, there is no model function that can be associated to the metric functions. If $F$ is, for instance, periodic, it strictly does not have an inverse as well, but we can manage to limit the domain in such a way as to make both $F$ and $F^{-1}$ well-defined. 

The $tt-$ and $rr-$ parts of the Einstein equation can be written respectively as
\begin{equation}
\label{eq:einstein_tt}
\begin{split}
&-4-2 r^2 L+4 r f' \\
&+ f \left[ 4 + r^2 \phi^{\prime 2} \left( 1 + G_X \left( f^\prime \phi^\prime + 2 f \phi^{\prime \prime}  \right) \right)  \right] = 0 
\end{split}
\end{equation}
\begin{equation}
\label{eq:einstein_rr}
\begin{split}
&-4 -  2 r^2 L + 4 f + 4rf\frac{h^\prime}{h} \\
&- r^2 f \phi^{\prime 2} + r \frac{f^2}{h} G_X \left( 4h + rh^\prime \right) \phi^{\prime 3} = 0.
\end{split}
\end{equation}
Eqs.~(\ref{eq:einstein_tt}), (\ref{eq:einstein_rr}), and (\ref{eq:current}) are derived explicitly in the Appendix. Using Eq.~(\ref{eq:current_constraint_1}) to remove the model dependence in the $tt$-component of the Einstein equation, we obtain
\begin{equation}
\label{eq:necessary_tt_Jr}
\begin{split}
f \phi^\prime \phi^{\prime \prime} + \frac{1}{2} f^\prime \phi^{\prime 2} = & \left( \frac{4}{r} + \frac{h^\prime}{h} \right) \\
& \times \left( - \frac{f^\prime}{r} - \frac{f}{r^2} - \frac{f}{4} \phi^{\prime 2} + \frac{L}{2} + \frac{1}{r^2} \right) .
\end{split}
\end{equation}
Similarly, using Eq. (\ref{eq:current_constraint_1}) in the $rr$-component of the Einstein equation, we obtain
\begin{equation}
\label{eq:necessary_rr_Jr}
\phi^{\prime 2} = \frac{2L}{f} + \frac{4}{r^2 f} - \frac{4 h^\prime}{h r} - \frac{4}{r^2} .
\end{equation}
Eqs. (\ref{eq:necessary_tt_Jr}) and (\ref{eq:necessary_rr_Jr}) are necessary conditions for any solution. They are obviously independent of the choice of $G(X)$, and in the space where radial component of the current vanishes, they are equivalent to the Einstein equation components. Manipulating Eq. (\ref{eq:necessary_rr_Jr}), we can write an expression for the kinetic density $\left( X = -f {\phi'}^2/2 \right)$ in terms of the metric functions:
\begin{equation}
\label{eq:necessary_constaint_X_1}
- 2 \frac{X}{f} =  \frac{2L}{f} + \frac{4}{r^2 f} - \frac{4 h^\prime}{h r} - \frac{4}{r^2} .
\end{equation} 
Later on, when we obtain explicit expressions for the metric functions in terms of $r$, we will match Eqs. (\ref{eq:necessary_constaint_X_1}) and (\ref{eq:prayer_function}) to construct the function $F$, which will then provide the model function $G_{X} = 1/\sqrt{-X} F(X)$ that supports a hairy solution. 

To find the metric functions, we use Eq. (\ref{eq:necessary_rr_Jr}) to eliminate the remaining $\phi^\prime$ dependence in Eq. (\ref{eq:necessary_tt_Jr}). The result is
\begin{equation}
\label{eq:necessary_tt_rr_Jr_1}
-\frac{4}{r^3} - 2\frac{f}{r} \frac{h^{\prime\prime}}{h} + \frac{f}{h^2} \frac{h^{\prime 2}}{r} - 2 \frac{f}{h} \frac{h^\prime}{r^2} + \frac{4 f}{r^3} - \frac{f^\prime h^\prime}{h r} + 2 \frac{f^\prime}{r^2}  = 0 .
\end{equation}
Eq. (\ref{eq:necessary_tt_rr_Jr_1}) is a necessary condition obtained by a combination of field equations and is equivalent to the $tt$-component of the Einstein equation in the space where $J^r = 0$ and the $rr$-component of the Einstein equation is satisfied. 

It is convenient to express the necessary conditions in terms of the logarithmic derivative
\begin{equation}
\label{eq:H_definition}
q(r) = \frac{d}{dr} \ln h(r)
\end{equation}
since it is only $q$ and its first derivative that appears in the equations. With this we can consider Eq. $(\ref{eq:necessary_tt_rr_Jr_1})$ as an expression that gives the functional $f[q]$ or $q[f]$. The necessary conditions can then be summarized in the form:
\begin{equation}
\label{eq:necessary_tt_rr_Jr}
f' + \left(\dfrac{4- 2rq-r^2(q^2+2 q')}{r(2-rq)}\right)f = \dfrac{4}{r(2-rq)} ,
\end{equation}
\begin{equation}
\label{eq:necessary_constraint_X}
X[q,f] = -L - \dfrac{2}{r^2} + \dfrac{2f[q]}{r}\left(q[f]+\dfrac{1}{r}\right) ,
\end{equation} 
\begin{equation}
\label{eq:current_constraint}
G_X \sqrt{-X} \sqrt{\frac{f[q]}{2}} \left( q[f]+ \frac{4}{r} \right) = 1 .
\end{equation}
What makes these equations convenient are the following characteristics: (a) Eq. (\ref{eq:necessary_tt_rr_Jr}) is completely model-independent (i.e. there is no $G_X$) and decoupled from the scalar field, and (b) Eq. (\ref{eq:necessary_constraint_X}) readily gives the kinetic density (and thus the scalar field) in terms of the metric functions. Eq. (\ref{eq:necessary_tt_rr_Jr}) is a linear equation for $f$ and a Riccati equation for $q$. It determines $f$ given $q$ (i.e. $f[q]$) or $q$ given $f$ (i.e. $q[f]$), and can be taken as the defining functional relation between the two metric functions. It is also valid in GR. One way to see this is to substitute the Schwarzschild-de Sitter solution into Eq. (\ref{eq:necessary_tt_rr_Jr}).  In the next section, this will also be made more transparent, as we show that if the metric functions are proportional to each other then the general solution to Eq. (\ref{eq:necessary_tt_rr_Jr}) has the Schwarzschild-de Sitter form. 

In what follows, we shall show how these equations can be put to good use. We shall use them to prove an extension of the no-hair theorem, and then to develop a method for obtaining static and spherically-symmetric hairy solutions in cubic Horndeski.

\subsection{Extending the no-hair theorem}
\label{subsec:no_hair_extension}

Consider, as a special case of Eq. (\ref{eq:necessary_tt_rr_Jr}), the case $h(r) = \kappa f(r)$ where $\kappa$ is a constant. This case includes GR-like black holes ($\kappa = 1$), i.e. stealth solutions, which we want to equip with scalar hair. Setting $h(r) = \kappa f(r)$ in Eq. (\ref{eq:necessary_tt_rr_Jr}) leads to the simple differential equation
\begin{equation}
\label{eq:necessary_tt_rr_Jr_special}
f'' - \frac{2}{r^2} \left( f - 1 \right) = 0 
\end{equation}
whose general solution is of the Schwarzschild-de Sitter form
\begin{equation}
\label{eq:SdS_form}
f(r) = 1 - \frac{M}{r} + \Lambda r^2
\end{equation}
where $M$ and $\Lambda$ are integration constants. Therefore, the most that we can hope for in a hairy, static and spherically-symmetric, solution with $h(r) = \kappa f(r)$ is a Schwarzschild-de Sitter (SdS) spacetime. We can push the results further by substituting the SdS metric function (Eq. (\ref{eq:SdS_form})) into the field equations given by Eqs. (\ref{eq:necessary_tt_Jr}) and (\ref{eq:necessary_rr_Jr}). 
In doing this and solving for the scalar field, we find that either equations require the scalar field to be given by 
\begin{equation}
\label{eq:scalar_hair}
\phi^\prime(r) = \sqrt{2 (L-6 \Lambda) } \sqrt{\frac{r}{-M+\Lambda  r^3+r}} .
\end{equation}
Since we are working with model-independent necessary conditions, Eq. (\ref{eq:scalar_hair}) has to be the only scalar hair that the cubic shift-symmetric Horndeski theory with $h(r) = \kappa f(r)$ can possess. We can therefore exclude the possibility of GR's SdS geometry $(L = 6 \Lambda)$ possessing scalar hair because this limit automatically gives a trivial scalar by Eq. (\ref{eq:scalar_hair}). The remaining condition, Eq. (\ref{eq:current_constraint}), requires Eq.~(\ref{eq:prayer_function}). This means that expressions for the kinetic density given by 
Eqs. (\ref{eq:prayer_function}) and (\ref{eq:necessary_constraint_X}) must be consistent. However, matching these two gives
\begin{equation}
F^{-1} \left( \frac{-3 M+6 \Lambda  r^3+4 r}{ \sqrt{2} r^2 \sqrt{1 -\frac{ M}{r}+ \Lambda  r^2}} \right) = 6 \Lambda - L,
\end{equation} 
which means that we require the existence of a function $F$ that maps a constant to a function of $r$. No such well-behaved function exists, and so there is no function $F$ (hence model function $G$) that can make all of the field equations consistent with each other. This is an extension to the no-hair theorem. To wit: among solutions of vanishing scalar current in the cubic sector of the shift-symmetric Horndeski gravity (which are not ruled out by the current no-hair theorems), there are still no static and spherically-symmetric hairy solutions of the form in Eq. (\ref{eq:metric_static_spherical}) with $h(r) = \kappa f(r)$. This class of solutions includes hairy stealth black holes. Hairy black hole solutions with a vanishing scalar current, should they exist, must therefore not be of the form $h(r) = \kappa f(r)$. In order to find them, we return to the coupled differential equations given in Eqs. (\ref{eq:necessary_tt_rr_Jr}), (\ref{eq:necessary_constraint_X}), and (\ref{eq:current_constraint}). \footnote{In Ref. \cite{st_horndeski_bhs_tatersall2018}, the spacetime represented by Eq. (19) in the text is claimed to be a black hole solution in the shift-symmetric cubic Horndeski theory defined by Eq. (\ref{eq:theory}). The solution should therefore satisfy the GR contraint Eq. (\ref{eq:necessary_tt_rr_Jr}). Unfortunately, we have been unable to verify this.}

\section{Some solutions in cubic Horndeski}
\label{sec:solutions_cubic_horndeski}

As shown in the previous section, if the metric functions are proportional to each other, then the field equations are inconsistent. Therefore, no model function, $G$, of cubic Horndeski can support the metric functions and scalar hair of that particular form. Outside this rather restrictive regime, we demonstrate in this section how to construct metrics able to support scalar hair, again using Eqs. (\ref{eq:necessary_tt_rr_Jr}), (\ref{eq:necessary_constraint_X}), and (\ref{eq:current_constraint}). We first demonstrate the key steps for three simple cases (sections \ref{subsec:h_constant}, \ref{subsec:f_constant}, and \ref{subsec:almost_black_hole_again}) before summarizing our general strategy for constructing hairy solutions (section \ref{subsec:recipe}). We discuss the method and in the next section reformulate it to better monitor curvature singularities that generically appear in the constructed solutions.

\subsection{Ultrastatic spacetimes: $h(r) =$ constant}
\label{subsec:h_constant}

Consider the case $h(r)$ is a constant so that the field equations (Eqs. (\ref{eq:necessary_tt_rr_Jr}), (\ref{eq:necessary_constraint_X}), and (\ref{eq:current_constraint})) reduce to
\begin{gather}
\label{eq:toy_1} G_X \sqrt{-X} \sqrt{\frac{f}{2}} \left(\frac{4}{r}\right) = 1 \\
\label{eq:toy_2} X = - L - \frac{2}{r^2} + \frac{2f}{r^2} \\
\label{eq:toy_3} f' + \left(\frac{2}{r}\right) f = \frac{2}{r} .
\end{gather}
The last equation is already uncoupled from the first two. This leads to the metric function
\begin{equation}
\label{eq:metric_function_toy}
f(r) = 1 - \frac{Q}{r^2} 
\end{equation}
where $Q$ is an integration constant. Substituting Eq. (\ref{eq:metric_function_toy}) into Eq. (\ref{eq:toy_2}) leads to
\begin{equation}
\label{eq:toy_2_upgrade} X = -L - \frac{2Q}{r^4} .
\end{equation}
The non-kinetic density factors in Eq. (\ref{eq:toy_1}) must be a function $F$ of the kinetic density only and that 
\begin{equation}
 \frac{\sqrt{8}}{r} \sqrt{1 - \frac{Q}{r^2}} = F \left( -L - \frac{2 Q}{r^4} \right) .
\end{equation}
This function $F$ exists and we can invert it because the right hand side is not a constant. It is easy to show that 
\begin{equation}
F(X) = 2\sqrt{ L + X + \sqrt{- \frac{2}{Q} } \sqrt{L + X} } .
\end{equation}
The trick is to consider Eq. (\ref{eq:toy_2_upgrade}) as an expression for the field $X$ in terms of the coordinate $r$. Now, we have a solution to cubic Horndeski with the model
\begin{equation}
\label{eq:cubic_galileon_solution_1}
G_X =\left( 2 \sqrt{-X} \sqrt{ L + X + \sqrt{- \frac{2}{Q} } \sqrt{L + X} } \right)^{-1} .
\end{equation}
By integrating the above result with respect to $X$, we then obtain the cubic model function with the metric function given by Eq. (\ref{eq:metric_function_toy}), a constant $h(r)$, and a nontrivial scalar field given by Eq. (\ref{eq:toy_2_upgrade}). It is easy to show that the solution represented by Eqs. (\ref{eq:cubic_galileon_solution_1}), (\ref{eq:toy_2_upgrade}), and (\ref{eq:metric_function_toy}) satisfy the field equations given by Eqs. (\ref{eq:einstein_tt}), (\ref{eq:einstein_rr}), and $J^r = 0$ where $J^r$ is given by Eq. (\ref{eq:current}). Because we are working with necessary conditions, our calculation here already exhausts all possible solutions with constant $h(r)$ and produces the only model that can support the scalar hair given by Eq. (\ref{eq:toy_2_upgrade}). 

Being ultrastatic, the norm of the timelike Killing vector, $\partial/\partial t$, does not vanish anywhere, and so there is no Killing horizon. The metric cannot be that of a black hole.
The Ricci scalar for the spacetime, $-2Q/r^4$, shows an essential singularity at the origin. The spacetime is asymptotically Ricci-flat while the scalar field is, in general, nontrivial at infinity, $\phi^{\prime 2} \sim L$ as $r \rightarrow \infty$. 

\subsection{$f(r) = $ constant}
\label{subsec:f_constant}

Another easy example is $f(r)$ being a constant. We assume that $f(r) = 1$ so that the field equations become
\begin{gather}
G_X \sqrt{- \frac{ X }{2} } \left( q + \frac{4}{r} \right) = 1 \\
X = -L + \frac{2 q}{r} \\
q^{\prime} + \frac{q}{r} = - \frac{q^2}{2} .
\end{gather}
The metric function $h$ is then given by 
\begin{equation}
\label{eq:metric_toy_h}
h(r) = a \left( 2b + \ln r \right)^2
\end{equation}
where $a$ and $b$ are integration constants and the kinetic density $X$ is given by
\begin{equation}
\label{eq:hair_toy_h}
X = -L + \frac{4}{r^2} \frac{1}{ 2b + \ln r }.
\end{equation}
This last equation readily gives the scalar hair. 
Once this equation is inverted for $r(X)$, the cubic model function possessing the metric function and scalar hair of Eqs. (\ref{eq:metric_toy_h}) and (\ref{eq:hair_toy_h}), respectively, is then given by
\begin{equation}
\label{eq:model_toy_h}
G_X  \frac{\sqrt{ - 2X }}{r(X)} \left( \frac{1}{2b + \ln r(X)} + 2 \right) = 1 .
\end{equation}  
It turns out that we can invert the kinetic density expression above semi-analytically using the Lambert W function which is the solution $x(y) = \text{W}(y)$ to the transcendental equation $x \exp x = y$. It is easy to show that
\begin{equation}
r = \sqrt{ \frac{8/(X+L)}{ \text{W} \left( 8 \exp \left(4b\right) / (X+L) \right) } } .
\end{equation}
The cubic model function possessing the metric function and the scalar hair given by Eqs. (\ref{eq:metric_toy_h}) and (\ref{eq:hair_toy_h}), respectively, is given by
\begin{equation}
\label{eq:cubic_galileon_toy_h}
\begin{split}
G_X = \frac{1}{4} \sqrt{ -\frac{2}{X} } &  \sqrt{ \frac{8/(X+L)}{ \text{W} \left( 8 \exp \left(4b\right) / (X+L) \right) } } \\ 
 & \times \left[ \frac{ \text{W} \left( 8 \exp \left(4b\right) / (X+L) \right) }{ \text{W} \left( 8 \exp \left(4b\right) / (X+L) \right)  + 1 } \right] .
\end{split}
\end{equation}

A notable feature here is that there are two integration constants compared to the only one appearing in the previous case. This is because when $h$ is supplied to decouple the field equations then Eq. (\ref{eq:necessary_tt_rr_Jr}) reduces to a first-order differential equation in $f$ but when $f$ is supplied then Eq. (\ref{eq:necessary_tt_rr_Jr}) reduces to a second-order differential equation in $h$. The solution is obviously not a black hole geometry because constant-$r$ hypersurfaces are not null. Nonetheless, an interesting limit of this solution is that when $a = 0$, the theory still possesses scalar hair even though its number of dimensions is effectively reduced and the space becomes Euclidean. Obviously, neither the geometry ($h(r) \sim a \left( \ln r \right)^2$) nor the scalar hair $\phi^{\prime 2} \sim L$ is asymptotically flat. The Ricci scalar $-2/(r^2 \left( 2b + \ln r \right))$ shows that the geometry contains an essential singularity at $r = 0$ and another at $r = \exp \left( -2b \right)$. The two singularities coincide only in the limit $b \rightarrow \infty$.

\subsection{$q^2 = - 2 q'$}
\label{subsec:almost_black_hole_again}

\begin{figure}[h]
\center
\includegraphics[width = 0.5 \textwidth]{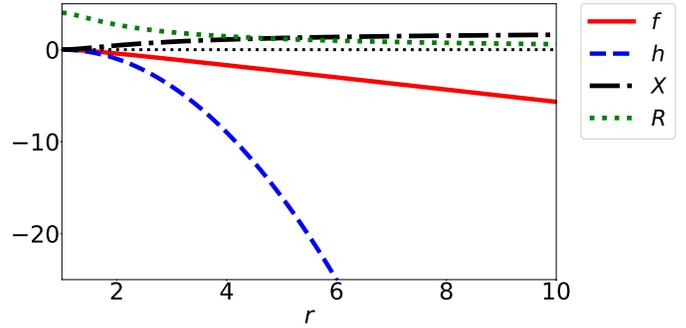}
\caption{Behaviour of the metric functions (Eqs. (\ref{eq:bv_bh_2_almost_h}) and (\ref{eq:bv_bh_2_almost_f})) and the corresponding Ricci scalar and the kinetic density in the case $L = -2$, $a = 1$, $b = -1$, and $c = - 1/3$.}
\label{fig:black_hole_solution_2}
\end{figure}

\begin{figure}[h]
\center
	\subfigure[ ]{
		\includegraphics[width = 0.48 \textwidth]{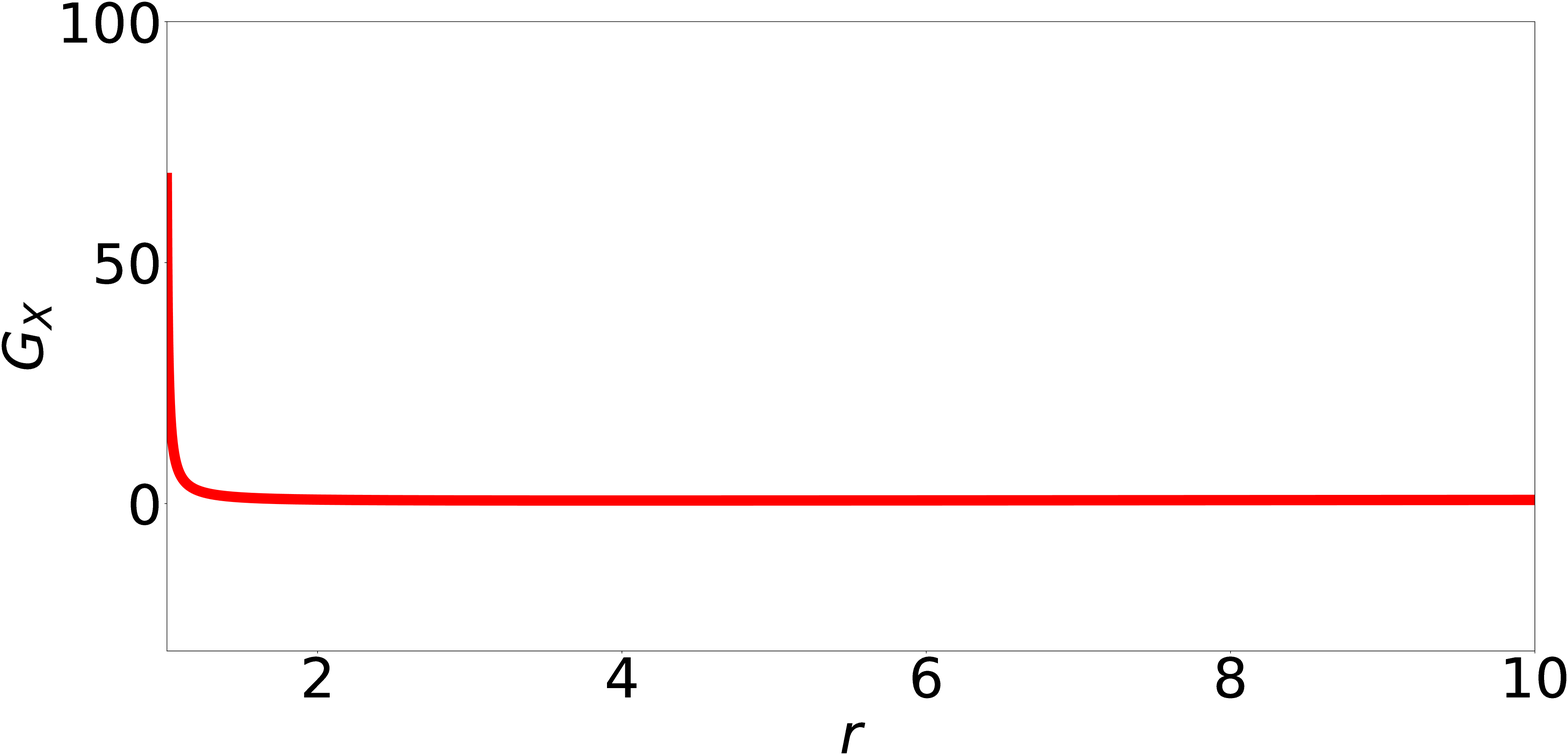}
		}
	\subfigure[ ]{
		\includegraphics[width = 0.48 \textwidth]{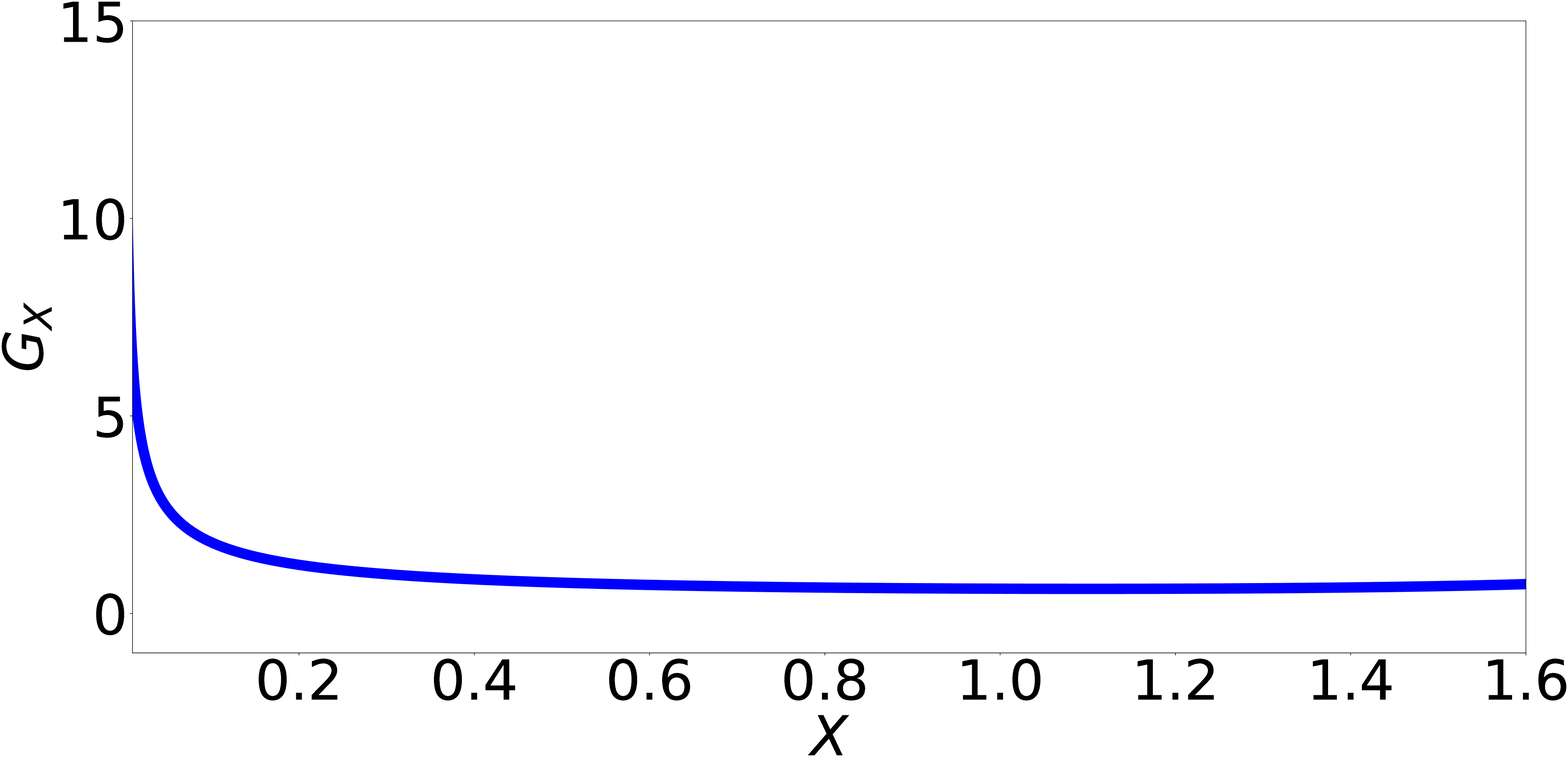}
		}
\caption{Model function $G_X$ as a function of the radial coordinate $r$ and the kinetic density $X$ for the solution given by Eqs. (\ref{eq:bv_bh_2_almost_h}) and (\ref{eq:bv_bh_2_almost_f}) in the case $L = -2$, $a = 1$, $b = -1$, and $c = - 1/3$. $G_X$ is finite both for $r=1$ and $X=0$.}
\label{fig:model_function_bv_bh_2}
\end{figure}

Another simple case that almost leads to a black hole solution is when
\begin{equation}
q' = -\frac{1}{2} q^2 .
\end{equation}
In this case, the necessary conditions become
\begin{gather}
G_X \sqrt{-X} \sqrt{ \frac{f[q]}{2} } \left( q + \frac{4}{r} \right) = 1 \\
X =  -L - \dfrac{2}{r^2} + \dfrac{2f[q]}{r}\left(q+\dfrac{1}{r}\right) \\
f' + \frac{2}{r} f = \frac{4}{r \left( 2 - qr \right)} .
\end{gather}
The solution to $q$ is given by
\begin{equation}
q(r) = 2/(r - a)
\end{equation}
and from this it is easy to show that
\begin{equation}
\label{eq:bv_bh_2_almost_h}	h(r) = b \left( r - a \right)^2
\end{equation} 
and
\begin{equation}
\label{eq:bv_bh_2_almost_f}	   f(r) = 1 - \frac{2}{3} \frac{r}{a} + \frac{c}{r^2} 
\end{equation}
where $a$, $b$, and $c$ are integration constants. Figure \ref{fig:black_hole_solution_2} shows that the metric functions can vanish at the same coordinate $r = a$; however, the spacetime cannot represent a black hole since both metric functions $h$ and $f$ are negative for the domain $r > 1$. The coordinates $r$ and $t$ in Eq. (\ref{eq:metric_static_spherical}) are then time and radial coordinates, respectively.
Instead of the metric functions vanishing for a certain radial coordinate as in a black hole, the metric functions here vanishes for a certain time coordinate. The case where both metric functions $h$ and $f$ vanish at the same time coordinate can easily be isolated leading to
\begin{eqnarray}
z &=& r/a \\
\label{eq:bv2_h} h(z) &=& b a^2 \left( z - 1 \right)^2 \\
\label{eq:bv2_f} f(z) &=& 1 - \frac{2z}{3} - \frac{1}{3 z^2}  \\
\label{eq:bv2_X} X(z) &=& -L - \frac{2}{a^2 z^2}  \nonumber \\
& & + \frac{2}{z a^2} \left( 1 - \frac{2 z}{3} - \frac{1}{3 z^2} \right) \left( \frac{1}{z} + \frac{2}{z - 1} \right)
\end{eqnarray}
where $X$ defines the associated scalar hair. The kinetic density $X$ given by Eq. (\ref{eq:bv2_X}) can, in fact, be inverted analytically so it is possible to explicitly write down the cubic model function. However, this expression turns out to be exceedingly long and rather uninformative \footnote{The hair-dressing method typically leads to unwieldy analytical expressions for the model function and, in general, it is only possible to express the model function numerically.} so we instead plot the model function for a specific choice of parameters in Figure \ref{fig:model_function_bv_bh_2}.

\subsection{Recipe for hairy solutions, and black hole with a naked singularity}
\label{subsec:recipe}

\begin{figure}[h]
\center
\includegraphics[width = 0.50 \textwidth]{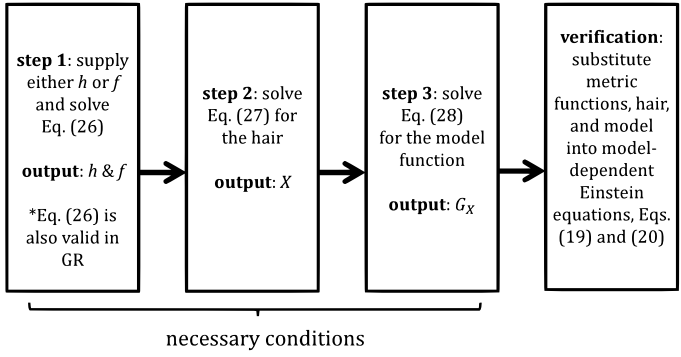}
\caption{Illustrative outline of the steps in our method.} 
\label{fig:recipe}
\end{figure}

The three foregoing examples flesh out a general recipe for obtaining hairy solutions in cubic Horndeski that we shall now summarize. We also illustrate the steps of our approach in Figure \ref{fig:recipe}. In summarizing, we also work out another example that ends up looking like an asymptotically-flat black hole, but that unfortunately is not regular outside its event horizon.  
\begin{enumerate}
\item Assume an ansatz that decouples the metric functions so that Eq. (\ref{eq:necessary_tt_rr_Jr}) reduces to an ordinary differential equation. Solve the resulting differential equation.

Say, we consider the ansatz
\begin{equation}
\label{eq:prayer_hole_h}
h(r) = 1 - \frac{M}{r} .
\end{equation}
In this case, Eq. (\ref{eq:necessary_tt_rr_Jr}) becomes a differential equation for $f(r)$ whose general solution is given by
\begin{equation}
\label{eq:prayer_hole_f}
f(r) = \frac{(M-r) \left(-M_0^2+12 M r-4 r^2\right)}{r (3 M-2 r)^2}
\end{equation}
where $M_0$ is an integration constant. Note that both $h(r)$ and $f(r)$ vanish at the radius $r = M$. 

\item Substitute the metric functions into Eq. (\ref{eq:necessary_constraint_X}) and invert the resulting expression for $r = r(X)$. If $X$ is not a constant, then the expression $X(r)$ represents the (derivative of the) scalar hair.

In the example considered in step 1, substituting the metric functions (Eqs. (\ref{eq:prayer_hole_h}) and (\ref{eq:prayer_hole_f})) into Eq. (\ref{eq:necessary_constraint_X}) leads to
\begin{equation}
\label{eq:prayer_hole_X}
X = \frac{2 M_0^2-9 M^2 \left(L r^2+2\right)+12 L M r^3-4 L r^4}{r^2 (3 M-2 r)^2}.
\end{equation}
This can be inverted to give
\begin{equation}
r(X) =\frac{1}{4} \left( 3M + \sqrt{ 9 M^2 + \frac{8 \sqrt{2} \sqrt{M_0^2-9 M^2}}{\sqrt{L+X}} } \right).
\end{equation}

\item Substitute $r(X)$ into Eq. (\ref{eq:current_constraint}) and solve for the derivative of the cubic model function $G_X$.

For the geometry defined by Eqs. (\ref{eq:prayer_hole_h}) and (\ref{eq:prayer_hole_f}) and the hair given by Eq. (\ref{eq:prayer_hole_X}), the model function can be explicitly written down if desired. Instead we end the example with the more useful implicit expression for the cubic model function:
\begin{equation}
\label{eq:generalized_cubic_galileon}
G_X (X) = \sqrt{ - \frac{2}{X f\left( r \left( X \right) \right) } } \left[ \frac{h^\prime \left( r \left( X \right) \right) }{ h \left( r \left( X \right) \right) } + \frac{4}{ r \left( X \right) } \right]^{-1}  .
\end{equation}

\end{enumerate}
Note that even if we cannot integrate $G_X = d G(X)/dX$ to obtain the actual model function $G(X)$ there is no real issue because it is $G_X$ that enters the scalar current and the stress-energy tensor of the scalar field. Finally, it is of course necessary to check if the results we obtain indeed satisfy the model-dependent starting field equations (Eqs. (\ref{eq:einstein_tt}) and (\ref{eq:einstein_rr}) and $J^r = 0$) and not just the reduced set of conditions that arose from combining them. Remarkably, all the results generated by our method so far have been checked to satisfy the original field equations. We illustrate the general approach in Figure \ref{fig:recipe}.

The geometry given by Eqs. (\ref{eq:prayer_hole_h}) and (\ref{eq:prayer_hole_f}) and the hair given by Eq. (\ref{eq:prayer_hole_X}) possesses some desirable properties. First, the geometry is asymptotically Schwarzschild,
\begin{eqnarray}
h(r) &=& 1 - \frac{M}{r} \\
f(r) &=& 1 - \frac{M}{r} + O \left(r^{-2}\right) .
\end{eqnarray}
Second, the hair is nontrivial even at infinity unless there is no cosmological constant,
\begin{equation}
\phi^{\prime 2} = 2L + O\left( r^{-4} \right) .
\end{equation}
The scalar hair, $\phi \sim 1/r$, drops as fast as the metric functions at infinity. Third, the singularity of the metric at $r = M$ is a coordinate singularity. One way to establish this is to go to a coordinate system that specializes to incoming light rays to probe what is behind the horizon $r = M$. In this case, we use the time coordinate
\begin{equation}
\label{eq:tortoise_time}
v = t + r^*(r)
\end{equation} 
where $r^*$ is defined by 
\begin{equation}
dr^* = \frac{dr}{ \sqrt{h(r) f(r)} } 
\end{equation}
so that the metric becomes 
\begin{equation}
\label{eq:metric_incoming_light_rays}
ds^2 = - h(r) dv^2 + 2 \sqrt{ \frac{h(r)}{f(r)} } dr dv + r^2 d\Omega^2 .
\end{equation}
Because both $h(r)$ and $f(r)$ vanish linearly at $r = M$, then the metric given by Eq. (\ref{eq:metric_incoming_light_rays}) is regular at $r = M$ and for all other $r > M$. This strengthens the claim that $r = M$ is a coordinate singularity and that we have in the spacetime given by Eqs. (\ref{eq:prayer_hole_h}) and (\ref{eq:prayer_hole_f}) a possible black hole with an event horizon at $r = M$. Furthermore, the singularity at $r = M$ does not manifest itself in the Ricci scalar which is given by
\begin{equation}
\label{eq:prayer_ricci}
R = \frac{4 \left(M_0^2-9 M^2\right)}{r (2 r-3 M)^3} .
\end{equation}
This expression features an essential singularity at $r = 0$. In addition, there is another essential singularity at $r=3M/2$ (which incidentally is the location of the light ring for a Schwarzschild black hole), outside the event horizon. This singularity also manifests itself in the metric given by Eq. (\ref{eq:metric_incoming_light_rays}) as the place where the geometry effectively becomes three-dimensional,
\begin{equation}
ds^2_{r=3M/2} = - h \left( \frac{3M}{2} \right) dv^2 + \left( \frac{3M}{2} \right)^2 d\Omega^2 .
\end{equation}
The spacetime then possesses a naked singularity, which is problematic for a number of reasons.
Attempting to add another adjustable parameter, say $Q$, by modifying the ansatz $h(r)$ to $\bar{h}(r) = h(r) + Q/r^2$ does not resolve this issue as it turns out that there is no solution to the equation which sets the essential singularity behind the event horizon. It is natural to question whether this unwanted feature is a consequence of the method. The essential singularity at $r \neq 0$, in the case discussed here, comes from $f(r)$ which is a solution to Eq. (\ref{eq:necessary_tt_rr_Jr}). So, for general static and spherically-symmetric solutions $h(r)$ and $f(r)$, we compute the Ricci scalar in the space where the constraint given by Eq. (\ref{eq:necessary_tt_rr_Jr}) holds. The resulting expression can be written as
\begin{equation}
\label{eq:ricci_necessary}
R = \frac{4}{r^2} - \frac{4f}{r^2} - \frac{3 f^\prime}{r} - \frac{f h^\prime}{r h} .
\end{equation}
Because the singularity at $r = 3M/2$ appears in the denominator of $f(r)$ given by Eq. (\ref{eq:prayer_hole_f}), the unwanted feature that we find here comes from the terms in Eq. (\ref{eq:ricci_necessary}) that are proportional to the metric function $f(r)$ and its derivative. This is a place where not only the Ricci scalar diverges but also the stress-energy tensor of the scalar field (Eqs. (\ref{eq:stress_energy_scalar_tt}) and (\ref{eq:stress_energy_scalar_rr})).

\section{Ricci scalar as dependent variable}
\label{sec:recipe_regular}

\begin{figure}[h]
\center
\includegraphics[width = 0.5 \textwidth]{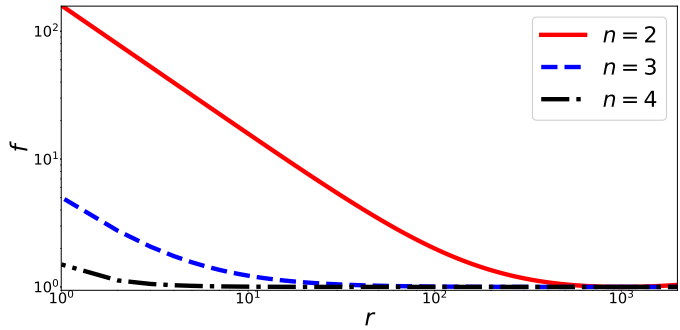}
\caption{Numerical solution of Eq. (\ref{eq:boss_battle_ode}) for $n = 2, 3, 4$ with initial condition $f = 1$ and $f' = 0$ at $r = 1000$.}
\label{fig:search_regular_bh}
\end{figure}

From our foregoing example, we can glean that even though it is easy to generate a solution to the field equations and detemine the cubic Horndeski model that supports it, the task of finding regular black hole solutions remains quite a challenge. Here, we slightly modify the field equations by adopting the Ricci scalar curvature as our dependent variable in lieu of the metric function $h$. Recognizing that only the ratio $h^\prime / h$ and its derivative appear in the field equations, we can use Eq. (\ref{eq:ricci_necessary}) to replace the metric function $h$ with the Ricci scalar $R$. With Eq.~(\ref{eq:H_definition}), Eq.~(\ref{eq:ricci_necessary}) takes the form
\begin{equation}
\label{eq:H_transformation}
q = -\frac{4}{r} + \frac{4}{r f} - 3 \frac{f^\prime}{f} - \frac{r R}{f} .
\end{equation}
Now, applying Eq.~(\ref{eq:H_transformation}) to Eq.~(\ref{eq:necessary_tt_rr_Jr}) leads to the following necessary condition involving the metric function $f$ and the Ricci scalar $R$:
\begin{equation}
\label{eq:necessary_tt_rr_Jr_revised}
\begin{split}
R' + \left( 8 - 4f - 7 r f^\prime \right) \frac{R}{2 r f} = & \frac{r}{2f} R^2 + \frac{1}{r^3 f} \bigg[ 8 - 14 f \\
& + 6 f^2 - 14 r f^\prime + 6 r f f^\prime \\
& + 6 r^2 f^{\prime 2} - 3 r^2 f^{\prime \prime} \bigg] .
\end{split}
\end{equation}
This can be considered as a first-order differential equation defining the functional $R[f]$ or a second-order differential equation defining the functional $f[R]$. The choice depends on how one wishes to start the process. Eq. (\ref{eq:necessary_tt_rr_Jr_revised}) is nothing but Eq. (\ref{eq:necessary_tt_rr_Jr}) only with the Ricci scalar $R$ instead of the metric function $h$. We can consider this as the starting point for the recipe and obtain the other metric function $h$ using Eqs. (\ref{eq:H_definition}) and (\ref{eq:H_transformation}). The advantage of this modified approach is that we can now start by specifying a desired \emph{regular} Ricci scalar, which was previously what blew up in previous examples. For instance, if we desire a geometry that is regular everywhere except at the origin, we can consider the simple choice
\begin{equation}
\label{eq:ricci_dream}
R = \frac{1}{r^n} .
\end{equation}
In this special case, Eq. (\ref{eq:necessary_tt_rr_Jr_revised}) turns to 
\begin{equation}
\label{eq:boss_battle_ode}
\begin{split}
& 6 r^2 f f^{\prime \prime} -7 r^{3-n} f^{\prime} \\
& -12 r^2 f^{\prime 2} -12 r f f^\prime +28 r f^\prime -2 n f r^{2-n}\\
&-4 f r^{2-n} -12 f^2 + 28 f + 8 r^{2-n}-r^{4-2 n}-16 = 0 .
\end{split}
\end{equation}
The solution of this non-linear second-order differential equation gives the metric function $f$ that is regular everywhere except at the origin. At this point, it is clear that the downside of controlling the singularities of the Ricci scalar better is having to deal with a considerably more difficult differential equation. One possible result of numerically integrating Eq. (\ref{eq:boss_battle_ode}) for some chosen values of $n$ is shown in Figure \ref{fig:search_regular_bh}. 
Immediately notable from this plot is that for this simple class of Ricci scalars, $f$ never vanishes, so we never get a black hole.
 
In any case, the foregoing example suggests an alternative version of general recipe which we present for completeness. We can replace step 1 of section \ref{subsec:recipe} by
\begin{enumerate}
\item Assume an ansatz that decouples the metric function $f$ and the Ricci scalar $R$ so that Eq. (\ref{eq:necessary_tt_rr_Jr_revised}) reduces to an ordinary differential equation. Solve the resulting differential equation and solve Eq. (\ref{eq:H_transformation}) for the metric function $h$.
\end{enumerate}
All the subsequent steps are the same, i.e. determine the scalar hair through Eq. (\ref{eq:necessary_constraint_X}), invert the kinetic density $X(r)$ to obtain $r(X)$, and solve for the model function given by Eq. (\ref{eq:generalized_cubic_galileon}).

\section{Black hole in cubic Horndeski}
\label{sec:black_hole}
 
\begin{figure}[h]
\center
\includegraphics[width = 0.5 \textwidth]{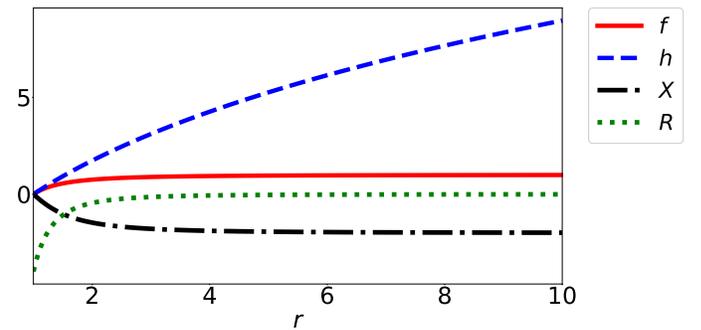}
\caption{Behaviour of the metric functions, the Ricci scalar, and the kinetic density given by Eqs. (\ref{eq:another_almost_f}), (\ref{eq:another_almost_h}), (\ref{eq:another_almost_R}), and \ref{eq:another_almost_X}) in the case $L = 2$, $a = 0$, and $Q = 1$.}
\label{fig:black_hole_solution}
\end{figure}

\begin{figure}[h]
\center
	\subfigure[ ]{
		\includegraphics[width = 0.48 \textwidth]{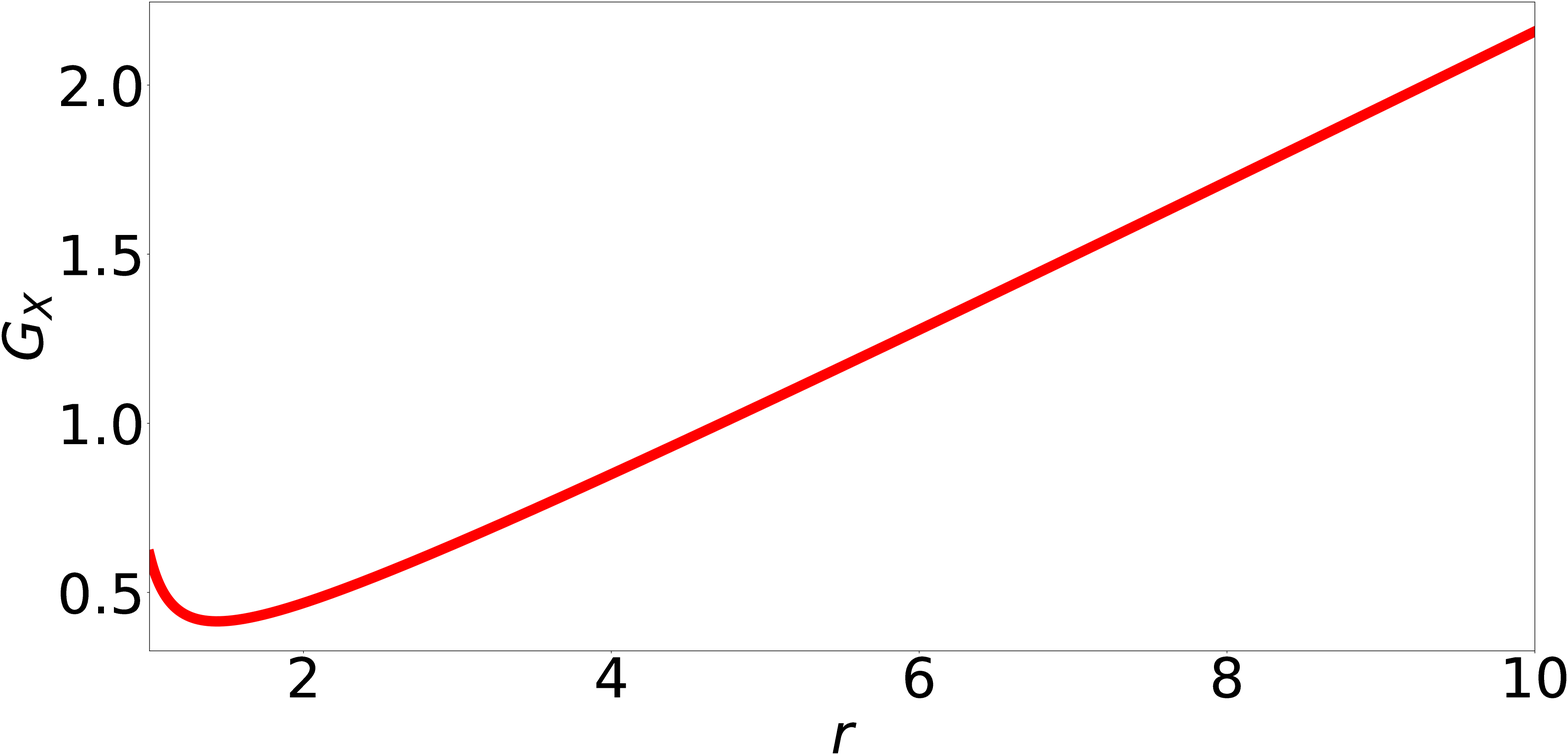}
		}
	\subfigure[ ]{
		\includegraphics[width = 0.48 \textwidth]{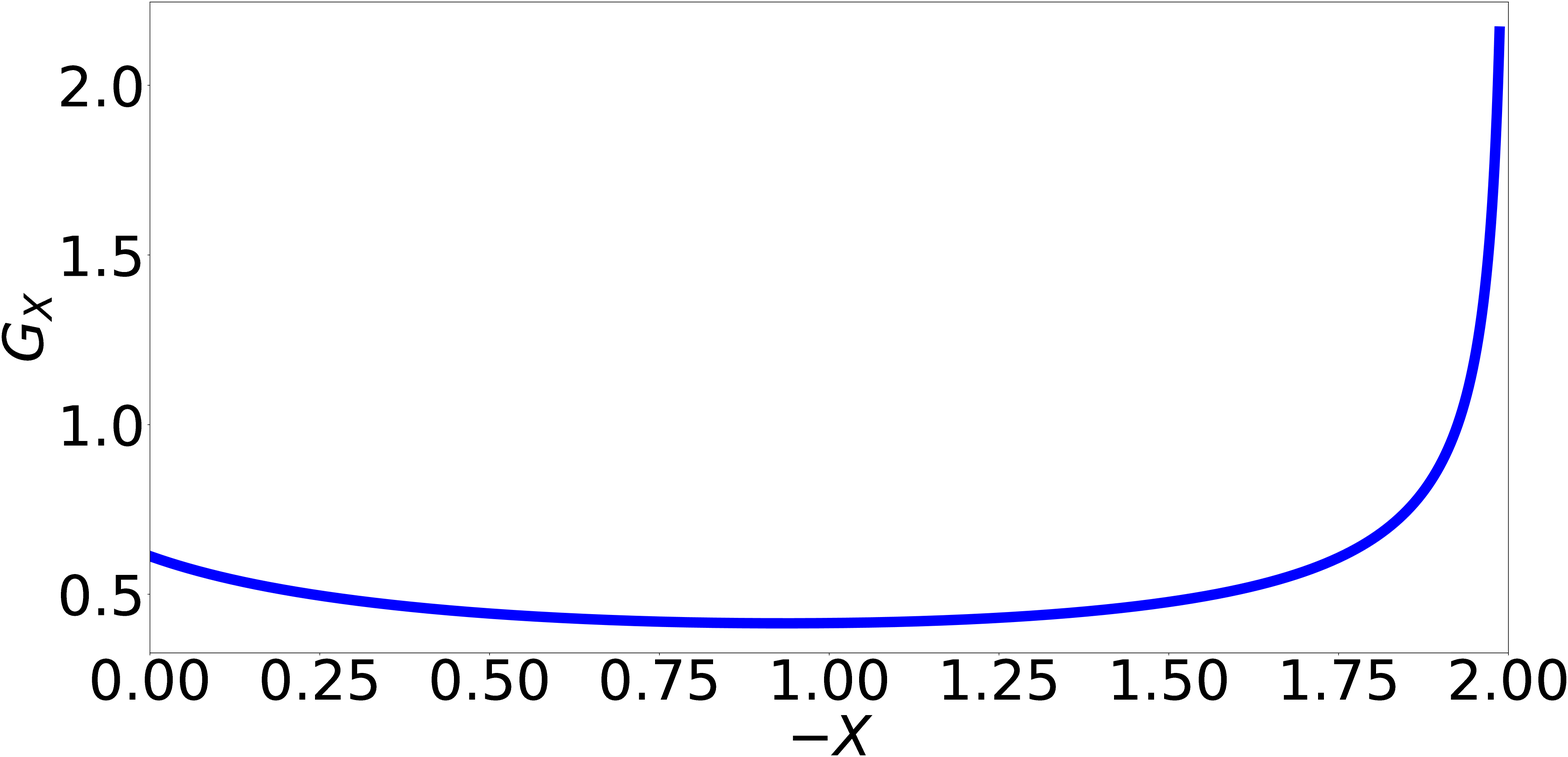}
		}
\caption{Model function $G_X$ as a function of the radial coordinate $r$ and the kinetic density $X$ for the black hole solution given by Eqs. (\ref{eq:another_almost_f}), (\ref{eq:another_almost_h}), and (\ref{eq:another_almost_X}) in the case $L = 2$, $a = 0$, and $Q = 1$.}
\label{fig:model_function_bh}
\end{figure}

\begin{figure}[h]
\center
\includegraphics[width = 0.5 \textwidth]{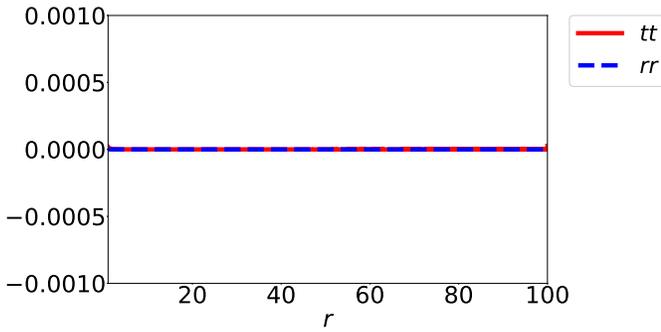}
\caption{Evaluated left-hand sides of Eqs. (\ref{eq:einstein_tt}) and (\ref{eq:einstein_rr}) -- the non-trivial components of the original field equations -- for the black hole described by Eqs. (\ref{eq:another_almost_f}) and (\ref{eq:another_almost_h}) with $L = 2$, $a = 0$, and $Q = 1$. This confirms that the result generated by the method is really a solution.}
\label{fig:field_equations_cubic_bh}
\end{figure}

A black hole solution can be obtained with the method (in either of its versions).  We begin by considering the metric function
\begin{equation}
\label{eq:another_almost_f}
f(r) = 1 - \frac{Q}{r^2} .
\end{equation}
With this ansatz, Eq. (\ref{eq:necessary_tt_rr_Jr_revised}) reduces to 
\begin{equation}
\left(2 r^3-2 Q r\right) R^\prime +\left(4 r^2-10 Q\right) R -\frac{8 Q}{r^2}+r^4 \left( -R^2 \right) = 0
\end{equation}
which is a nonlinear first-order differential equation in $R$. The general solution to this is given by 
\begin{equation}
\label{eq:another_almost_R}
R\left(r\right) = -\frac{2 \left(a Q+r \sqrt{r^2-Q}+Q \ln \left(\sqrt{r^2-Q}+r\right)\right)}{r^4 \left(a+\ln \left(\sqrt{r^2-Q}+r\right)\right)} 
\end{equation}
where $a$ is an integration constant. Using the above expressions for $f$ and $R$, we then solve Eq. (\ref{eq:H_transformation}), a linear homogeneous first-order differential equation for $h$, to obtain
\begin{equation}
\label{eq:another_almost_h}
h\left(r\right) = b \left(a+\ln \left(\sqrt{r^2-Q}+r\right)\right)^2 
\end{equation}
where $b$ is an integration constant that can be set to unity with a redefinition of $t$. The above expressions for $h$ and $f$ satisfy Eq. (\ref{eq:necessary_tt_rr_Jr}) as expected. The scalar hair given by Eq. (\ref{eq:necessary_constraint_X}) becomes
\begin{equation}
\label{eq:another_almost_X}
X = \frac{2}{r^4} \left(\frac{2 r \sqrt{r^2-Q}}{a+\ln \left(\sqrt{r^2-Q}+r\right)}-Q\right)-L .
\end{equation}
Figure \ref{fig:black_hole_solution} shows that the parameters can be selected to give a black hole spacetime.  The metric functions can be made to vanish at $r=\sqrt{Q}$ (i.e. the putative event horizon) with the same scaling:
\begin{equation}
f(r) \sim h(r) \sim 2 (r - \sqrt{Q}). 
\end{equation}
From this figure we also see that the spacetime is asymptotically Ricci-flat. The kinetic density $X$, on the other hand, approaches a constant value, which means that the overall solution is not asymptotically flat. Finally, the kinetic density is negative and not constant, so it can be numerically inverted to obtain the model function $G_X$. Figure \ref{fig:model_function_bh} shows the model function $G_X$ in terms of the radial position $r$ and the corresponding kinetic density $X$.
Substituting Eqs. (\ref{eq:another_almost_f}) and (\ref{eq:another_almost_h}) and the model function presented in Figure \ref{fig:model_function_bh} into the original Einstein field equations (Eqs. (\ref{eq:einstein_tt}) and (\ref{eq:einstein_rr})) leads to Figure \ref{fig:field_equations_cubic_bh}. This shows that the model-dependent field equations are indeed satisfied, and confirms our result as a Horndeski solution.

One intriguing feature of the solution is its unusual asymptotics. It presents a logarithmic divergence at infinity. This logarithmic divergence, nevertheless, is much slower compared to the parabolic divergence of the Schwarzschild-(anti) de Sitter black holes and the hairy black hole solutions in Horndeski theory with non-minimal derivative coupling to the Einstein tensor \cite{st_horndeski_solutions_kobayashi, st_horndeski_solutions_babichev_0, st_horndeski_solutions_rinaldi, st_horndeski_solutions_anabalon, st_horndeski_solutions_minamitsuji}. 

\begin{figure}[h]
\center
\includegraphics[width=0.5 \textwidth]{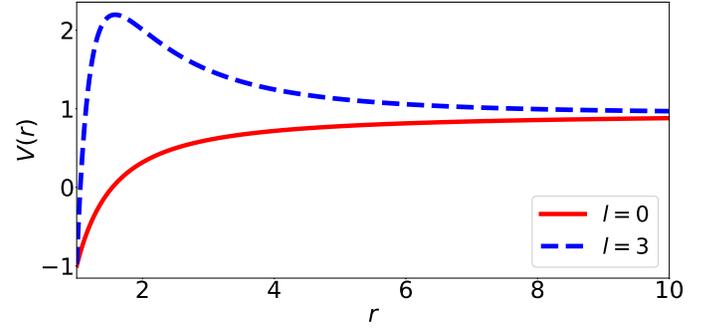}
\caption{Potential function (Eq. (\ref{eq:potential_function})) for radial ($l = 0$) and non-radial ($l=3$) geodesics for the black hole defined by Eqs. (\ref{eq:another_almost_f}) and (\ref{eq:another_almost_h}) for $a = 0$ and $Q = 1$.}
\label{fig:potential_cubic_bh}
\end{figure}

\begin{figure}[h]
\center
\includegraphics[width = 0.5 \textwidth]{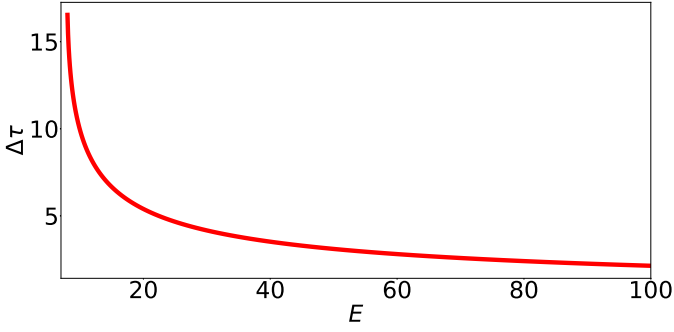}
\caption{Proper time to radially plunge from a radius $r = 10$ down to the horizon $r = 1$ as a function of the energy $E$ for the black hole described by Eqs. (\ref{eq:another_almost_f}) and (\ref{eq:another_almost_h}) for $a = 0$ and $Q = 1$.}
\label{fig:proper_time_plot}
\end{figure}

It is reasonably straightforward to show that the event horizon can be reached from any finite $r$ in finite proper time. Since we are in the Jordan frame where point particles move along the geodesics of the spacetime, the radial motion of point particles can be described by
\begin{equation}
\label{eq:newton_like_geodesics}
\left( \frac{dr}{d\tau} \right)^2 = \frac{f(r)}{h(r)} E - V(r) 
\end{equation}
where
\begin{equation}
\label{eq:potential_function}
V(r) = f(r) \left( 1 + \frac{l^2}{r^2} \right) - \frac{f(r)}{h(r)}
\end{equation}
and $E$ and $l$ are constants along the geodesics defined by
\begin{eqnarray}
\sqrt{E + 1} &=& h(r) \frac{dt}{d\tau} \\
l &=& r^2 \sin^2 \theta \frac{d\phi}{d\tau}
\end{eqnarray}
and $\tau$ is the proper time. For radial and non-radial geodesics the shape of the potential function is shown in Figure \ref{fig:potential_cubic_bh}.
The proper time $\Delta \tau$ it takes for an object to fall from an initial radial position $r_i$ down to the horizon $r_H$ is then given by
\begin{eqnarray}
\Delta \tau = - \int_{r_i}^{r_H} dr \frac{1}{ \sqrt{ \frac{f(r)}{h(r)} E - V(r) } }.
\end{eqnarray}
The integral above cannot be solved analytically for the spacetime given by Eqs. (\ref{eq:another_almost_f}) and (\ref{eq:another_almost_h}). Nonetheless, it can be integrated numerically. For instance, if the initial position is $r = 10$, $l=0$, and $E = 8$ then $\Delta \tau = 17$ and if $E = 10$ then $\Delta \tau = 10$. Generally, as the energy $E$ increases it takes less time for the object to fall down the horizon at $r = 1$. The proper time needed to reach the horizon as a function of the energy $E$ is plotted in Figure \ref{fig:proper_time_plot}. 

\begin{figure}[h]
\center
\includegraphics[width = 0.5 \textwidth]{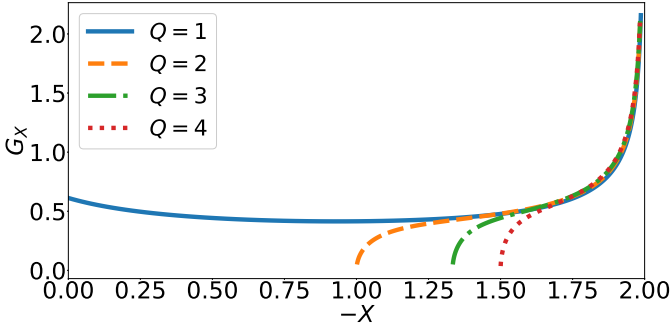}
\caption{Model function $G_X$ for the black hole described by Eqs. (\ref{eq:another_almost_f}) and (\ref{eq:another_almost_h}) for $a = - \ln \sqrt{Q}$ and $L = 2$.}
\label{fig:GX_q_dependence}
\end{figure}

One more aspect of the solution deserves comment: its parameter $Q$ determines the Horndeski model, $G_X$. The dependence of $G_X$ on $Q$ is displayed in Figure \ref{fig:GX_q_dependence}. The same can be said of all the other solutions we presented above, in that the model explicitly depends on the integration constants. The nature of this parameter is therefore similar to the Reissner-Norndstrom-like charge in Ref. \cite{st_horndeski_solutions_babichev_2} and for the black hole solutions in Horndeski theory with derivative coupling to the Einstein tensor \cite{st_horndeski_solutions_kobayashi, st_horndeski_solutions_babichev_0, st_horndeski_solutions_rinaldi, st_horndeski_solutions_anabalon, st_horndeski_solutions_minamitsuji, st_horndeski_solutions_gaete}; it represents secondary hair. 

\subsection{Hairy black hole in cubic Horndeski}

To select the black hole solutions of Eqs. (\ref{eq:another_almost_f}) and (\ref{eq:another_almost_h}) we require that $f$ and $h$ vanish at the coordinate $r = \sqrt{Q}$. The steps are easy and the resulting black hole solution can be written as
\begin{eqnarray}
f(x) &=& 1 - \frac{1}{x^2} \\
h(x) &=& \ln ^2 \left( \sqrt{ x^2 - 1  } + x \right) 
\end{eqnarray}
where
\begin{equation}
x = r / \sqrt{Q} .
\end{equation}
As previously mentioned, both metric functions behave like $\sim 2(x-1)$ as $x\rightarrow 1$. This makes $x=1$ a Killing horizon through which horizon-penetrating coordinates can be constructed in the standard manner. 
These metric functions imply that the kinetic density (which is essentially the scalar hair) and the Ricci scalar are, respectively, 
\begin{eqnarray}
X\left( x \right) &=& \frac{2}{Q} \frac{1}{x^3} \left[ \frac{2 \sqrt{ x^2 - 1 } }{ \ln \left( \sqrt{x^2 - 1} + x \right) } - \frac{1}{x} \right] - L \\
R\left( x \right) &=& -\frac{2}{Q} \frac{1}{x^4} \left[ 1 + \frac{ x \sqrt{ x^2 - 1 } }{ \ln \left( \sqrt{ x^2 - 1 } + x \right) }  \right] .
\end{eqnarray}
Both of these are regular functions for all $x >0$. The solution thus represents a regular, hairy black hole. Note that for $x>1$ (outside the horizon), the kinetic density $-X(x)$ ranges from $L-2/Q$ to $L$. We want this kinetic density to be negative, thus requiring the constraint $2 \leq QL$. Since $Q$ has to be positive for the spacetime to be a black hole, $L$ also has to be positive. This boundedness in $-X$ also occurs for the black hole solutions of Refs. \cite{st_horndeski_solutions_kobayashi, st_horndeski_solutions_babichev_0, st_horndeski_solutions_rinaldi, st_horndeski_solutions_anabalon, st_horndeski_solutions_minamitsuji}, though it may be more physically sensible for the theory to be well-defined for all values of the kinetic density $-X$, such as for instance, in the black hole solution of Ref. \cite{st_horndeski_solutions_babichev_2}. The singularities of the model function indicate where the theory breaks down. In Ref. \cite{st_horndeski_bhs_tatersall2018}, the theory $G_3 \left( X \right) = \alpha \ln \left( -X \right)$ is singular at $X = 0$, indicating the absence of an asymptotically flat limit. From our previous cases, the typical $G_X$ produced by our method possesses a singularity at $X = 0$. (Recall the results of Sections \ref{subsec:h_constant} and \ref{subsec:f_constant}, which might be considered as limits of the theory of Section \ref{sec:black_hole}). The typical singularity at $X = 0$ is perhaps not surprising, given that all these attempts need to evade the no-hair theorem of cubic Horndeski \cite{st_no_hair_theorem_hui, st_no_hair_benkel, st_no_hair_theorem_sotiriou_1, st_no_hair_theorem_sotiriou_2}. 

The model $G_X$ that supports this solution depends on $Q$. Since $G_X$ is implicitly defined implicitly in terms of $(f,h, X)$, we instead plot this in Figure \ref{fig:GX_q_dependence} for several choices of $Q$. In this figure, the singularity at $X = 0$ lies outside the allowed range of $-X$ for all but $Q=1$. On the other hand, when $Q=2/L$, the model can be shown to be finite at $X=0$. (More specifically, $G_X \sim \text{constant}$ as $X \rightarrow 0$ when $Q = 2/L$.) 

\subsection{Asymptotics at the horizon}
\label{subsec:asymptotics_horizon}

In the limit $x \rightarrow 1$, the solution behaves as
\begin{eqnarray}
f(x) &\sim & 2 (x - 1) - 3 (x - 1)^2 \\
h(x) &\sim & 2 (x-1)-\frac{1}{3} (x-1)^2 \\
X(x) &\sim & \left(\frac{2}{Q}-L\right)-\frac{8 (x-1)}{3 Q}-\frac{4 (x-1)^2}{45 Q} \\
R(x) &\sim & -\frac{4}{Q} +\frac{40 (x-1)}{3 Q} -\frac{1348 (x-1)^2}{45 Q} .
\end{eqnarray}
Clearly, the solution is a black hole as both the metric functions vanish at the horizon at the same rate. This also means that there is a coordinate system described by Eq. (\ref{eq:metric_incoming_light_rays}) where the singularity at the horizon disappears. The kinetic density approaches a constant value at the horizon meaning that the scalar field derivative $\phi^\prime$ diverges as $1/\sqrt{x - 1}$. More importantly, the formalism exploited here guarantees the finiteness of the scalar current at the horizon. It can also be shown that the scalar field itself vanishes as $\sqrt{x - 1}$ at the horizon. Figure \ref{fig:black_hole_solution} features the general shape of the kinetic density as a function of the coordinate $r$ and shows that the maximum value of the kinetic density is achieved at the horizon. Since the kinetic density must be negative, we may use the asymptotic value given above to show that $Q$ and $L$ should satisfy $QL > 2$. This defines the parameter space where the black hole solution can be associated with a real model function in cubic Horndeski. The Ricci scalar is also a constant at the horizon and with the previous constraint on $Q$ and $L$ it can be shown that the Ricci scalar at the horizon is related to the cosmological constant as $R < - 2L$. 

\subsection{Asymptotics at infinity}
\label{subsec:asymptotics_infinity}

In the limit $x \rightarrow \infty$, the solution behaves as
\begin{eqnarray}
f(x) &\sim & 1-\left(\frac{1}{x}\right)^2 \\
h(x) &\sim & \ln (2x) \left[ \ln (2x) -  \frac{1 }{ 2 x^2 } \right] \\
X(x) &\sim & -L + \frac{4}{Q x^2 \ln (2 x) } \\
R(x) &\sim & -\frac{2}{Q x^2 \ln (2 x) } .
\end{eqnarray}
We note the following properties. The spacetime is asymptotically Ricci-flat. If there is no cosmological constant, the scalar field would also be a constant. The constraint on $Q$ and $L$ to keep the kinetic density negative everywhere, however, prevents us from setting $L$ to zero so that, in general, the scalar field can only behave as $x$ asymptotically. The metric function $h$ logarithmically diverges at infinity, which is starkly different from the parabolic divergence (i.e. Schwarzschild-de Sitter) characterizing the asymptotics of other hairy black holes in Horndeski theory \cite{st_horndeski_solutions_kobayashi, st_horndeski_solutions_babichev_0, st_horndeski_solutions_rinaldi, st_horndeski_solutions_anabalon, st_horndeski_solutions_minamitsuji}. The asymptotic behavior at the horizon and at infinity and the monotonic behavior of the function $X(x)$ shows that the kinetic density is bounded below by $-L$ and above by $2/Q - L$. This boundedness of the kinetic density is also seen in solutions of theories with non-minimal derivative coupling to the Einstein tensor \cite{st_horndeski_solutions_kobayashi, st_horndeski_solutions_babichev_0, st_horndeski_solutions_rinaldi, st_horndeski_solutions_anabalon, st_horndeski_solutions_minamitsuji}.

\section{Discussion}
\label{sec:discussion}

The astute reader will probably notice that the same metric function $f = 1 - Q/r^2$ appears for both the black hole spacetime of Section \ref{sec:black_hole} and the ultrastatic spacetime of Section \ref{subsec:h_constant}. This points to a very interesting question: why does supplying the ultrastatic ansatz (constant $h$) lead to a spacetime with $f = 1- Q/r^2$, but then supplying $f = 1 - Q/r^2$ does not lead to an ultrastatic spacetime but instead to a black hole spacetime? From the discussion above, the ultrastatic branch seems to be inaccessible when $f = 1 - Q/r^2$ is fed into Eq. (\ref{eq:necessary_tt_rr_Jr}) as a starting point. For one, we comment that the spacetime of Section \ref{subsec:f_constant} is a limit of the black hole spacetime for $Q \rightarrow 0$ while the ultrastatic spacetime of Section \ref{subsec:h_constant} is only an improper limit of the black hole spacetime for $a \rightarrow \infty$. One might then think of the hairy black hole spacetime of Section \ref{sec:black_hole} as a generalization of the hairy (non-black hole) spacetimes of Sections \ref{subsec:h_constant} and \ref{subsec:f_constant}. But more importantly, this observation leads to an even more interesting question: is the hairy spacetime generated by the method, $\left( h, f, X \right)$, unique to a cubic Horndeski model? 
 
The resolution to this puzzle rests with recognizing that our method treats the metric functions $h$ and $f$ differently. 
In particular, the necessary condition given by Eq. (\ref{eq:necessary_tt_rr_Jr}) is a \emph{nonlinear} (Riccati) differential equation for $h$ and a \emph{linear} differential equation for $f$. This means that in assuming some $h$, one can uniquely determine $f$, fix the scalar hair $X$, and then associate to these a unique cubic Horndeski theory. But when starting with $f$, the resulting Ricatti equation can have many solutions for $h$. Each distinct pair $(f,h)$ then determines a unique theory $G_X$. This is precisely what happens above. When $f = 1 - Q/r^2$ is used as a starting point, the resulting Ricatti equation admits both $h$ = constant (ultrastatic) and the $h(x) = \ln ^2 \left( \sqrt{ x^2 - 1  } + x \right)$ (black-hole) as solutions. 

This distinction does not affect the scope of the no-hair extension of Section \ref{subsec:no_hair_extension}, because there Eq. (\ref{eq:necessary_tt_rr_Jr}) reduces to a simple linear differential equation when the metric functions are proportional up to a constant. This is also the case when the method is applied in a cosmological context, where the counterpart differential equation is also linear, and the method presents no ambiguity in the assignment of a solution (completely described by a scale factor $a$ and matter density $\rho$) to a cubic Horndeski theory \cite{st_horndeski_cosmological_solutions_bernardo}.

Finally, we remark that the theories obtainable using our method are of the second-type in the classification of Sotiriou and Zhou \cite{st_no_hair_theorem_sotiriou_1}, which is described by nonanalytic model functions\footnote{First-type theories fall under the no-hair class while the third-type theories are defined by a conformal coupling between the scalar field and the Gauss-Bonnet scalar \cite{st_no_hair_theorem_sotiriou_1}.}. In this class of theories, the scalar field is forced to put on a nontrivial profile and the solutions can potentially diverge at infinity due to the possible absense of Minkowski vacuum \cite{st_horndeski_solutions_kobayashi, st_horndeski_solutions_babichev_0, st_horndeski_solutions_rinaldi, st_horndeski_solutions_anabalon, st_horndeski_solutions_minamitsuji}.

\section{Concluding remarks}
\label{sec:conclusions}

We have shown that instead of the usual procedure of specifying the Horndeski model and solving for the metric functions and scalar field, as is done in Refs. \cite{st_horndeski_babichev, st_horndeski_solutions_babichev_2, st_horndeski_cosmological_tuning_babichev, st_horndeski_slow_rotation_bh_maselli, st_horndeski_neutron_stars_maselli, st_horndeski_solutions_kobayashi, st_horndeski_solutions_babichev_0, st_horndeski_solutions_rinaldi, st_horndeski_solutions_anabalon, st_horndeski_solutions_minamitsuji, st_horndeski_solutions_gaete}, we can reverse the process by (partially) starting from the metric functions, computing the corresponding scalar hair, and then finally determining the model in the cubic sector of shift-symmetric Horndeski that supports them. For static and spherically-symmetric scalar and tensor fields, one of the outputs of this technique happens to be a new hairy black hole (Section \ref{sec:black_hole}). The question of its stability remains unexplored, and we expect to return to this matter in future work. To the best of our knowledge, this is the first exact black hole solution in the phenomenologically-interesting cubic sector of Horndeski gravity that does not have an identical GR counterpart. (See, however, \cite{st_horndeski_solutions_minamitsuji_2} for hairy stealth solutions in this sector.) Another result of this reverse procedure is a theorem which states that there is no model function that can be assigned if the metric functions are proportional to each other and the radial component of the current is zero. This covers black holes such as that in GR where $h = f$ in the line element given by Eq. (\ref{eq:metric_static_spherical}). A hairy black hole solution with $h = f$ can still be built asymptotically with a time-dependent scalar field such as that in Ref. \cite{st_horndeski_cubic_babichev} for the cubic galileon theory.

To end, we discuss the scope and limitations of the method and its possible extension to include the other generalized galileon terms of Horndeski gravity. One limitation is that, in general, we can only obtain the model function $G(X)$ implicitly even if we can have analytical expressions for the metric functions and the scalar hair. This is because in many cases, we cannot do the inversion in the second step of the recipe analytically to obtain $r(X)$. Also, the galileon models arising from our method are obviously non-elementary functions of the kinetic density, brushing aside the issue that we are actually ending the process with $G_X = dG/dX$ and leaving the remaining integration implicit. A useful feature of the cubic shift-symmetric sector that allows us to bypass some of these issues is that only the derivative $G_X$ appears in the field equations. Another limitation is that we generally cannot associate resulting integration constants to Noether charges because the model function also depends on them. This is the case for many of the black hole solutions in Horndeski theory  \cite{st_horndeski_babichev, st_horndeski_solutions_babichev_2, st_horndeski_cosmological_tuning_babichev, st_horndeski_slow_rotation_bh_maselli, st_horndeski_neutron_stars_maselli, st_horndeski_solutions_kobayashi, st_horndeski_solutions_babichev_0, st_horndeski_solutions_rinaldi, st_horndeski_solutions_anabalon, st_horndeski_solutions_minamitsuji, st_horndeski_solutions_gaete} where the deviation of the metric from the GR solutions is fixed by the model parameters. Thus, we can at best hope for secondary hair using the method presented here. 

The method was simple to devise because there is only a single arbitrary function $G_X$ that appears in all of the field equations in cubic shift-symmetric Horndeski. This sector of Horndeski gravity is unique in this sense. As a starting point, we can therefore immediately eliminate the model dependencies in the field equations. A natural extension of the method is to include the other generalized galileon terms, say, a general quadratic model function, $G_2(X) = H(X)$. The issue here is the appearance of a combination of $G_2$ and $G_{2X}$ in the field equations that might require a different threading of the field equations to obtain model-independent necessary conditions. The situation is, of course, more challenging if we attempt to generalize the method to the quartic and quintic cases because even the second derivatives of the galileon model functions enter the field equations. The vanishing of the radial component of the current forces a constraint on the model functions that we found to be very restrictive in eliminating the model dependence of the field equations.  

The primary advantage of our method is that it gives explicit expressions for the metric functions and the spacetime hair. It can also, in fact, be used to design spacetimes for other purposes and not just black holes. This is a promising avenue for research that we intend to pursue with future work. As we intimated in the Introduction, the method is -- by fiat -- similar to that used for studying single-scalar field potential models of cosmology. In a follow-up paper, we show how it can be used to assign a cosmological scenario such as inflation to cubic Horndeski \cite{st_horndeski_cosmological_solutions_bernardo}.

\appendix*

\section{Field equations}
\label{sec:field_equations}

Here, we show the derivation of Eqs. (\ref{eq:current}), (\ref{eq:einstein_tt}), and (\ref{eq:einstein_rr}) from the covariant field equations represented by Eqs. (\ref{eq:covariant_einstein_equation}) and (\ref{eq:covariant_scalar_equation}) 
where the stress-energy tensor of the scalar field $T_{\alpha \beta}^{(\phi)}$ is given by \cite{st_horndeski_slow_rotation_bh_maselli, perturbation_kobayashi_1, perturbation_kobayashi_2} 
\begin{equation}
\label{eq:stress_energy}
\begin{split}
8 \pi T^{(\phi)}_{\alpha \beta} = &\frac{1}{2} G_2 g_{\alpha \beta} + \frac{1}{2} G_{2X} \left( \partial_\alpha \phi \right)\left( \partial_\beta \phi \right) \\
&+ \bigg[ -\frac{1}{2} G_{3X} \Box \phi \left( \partial_\alpha \phi \right) \left( \partial_\beta \phi \right) \\
& \ \ \ \ \ \ \ +\frac{1}{2} G_{3\mu} \left( \partial^\mu \phi \right) g_{\alpha \beta} - G_{3(\alpha} \left( \partial_{\beta )} \phi \right)  \bigg]  
\end{split}
\end{equation}
and the Noether current $J^\alpha(x)$ associated with the shift symmetry is given by 
\begin{equation}
\label{eq:current_covariant}
J^{\alpha}(x) = - \left( \partial^\alpha \phi \right) \left[ G_{2X} - G_{3X} \Box \phi \right] + G_{3X} \left( \partial^\alpha X \right) .
\end{equation}
Noting that 
\begin{eqnarray}
X &=& - \frac{1}{2} f ( \phi' )^2 \\
\partial_\alpha \phi &=& \delta^r_\alpha \phi^\prime
\end{eqnarray}
then we can write down the current as
\begin{equation}
\begin{split}
J^r 
&= - \left( \partial^r \phi \right) \left[ G_{2X} - G_{3X} \Box \phi \right] + G_{3X} \left( \partial^r X \right) 
\end{split}
\end{equation}
or simply
\begin{equation}
J^r = -f \phi' + \frac{f^2 (\phi')^2}{2} \frac{r h^{\prime} + 4 h}{r h} G_{X}
\end{equation}
which is the same as Eq. (\ref{eq:current}). The above stress-energy tensor can be written as
\begin{equation}
\begin{split}
8 \pi T^{(\phi)}_{\alpha \beta} =& \frac{1}{2} \left( X + L \right) g_{\alpha \beta} + \frac{1}{2} \delta^r_\alpha \delta^r_\beta \phi^{\prime 2} \\
& + G_X \bigg[ -\frac{1}{2} \Box \phi \delta^r_\alpha \delta^r_\beta \phi^{\prime 2} \\
& \ \ \ \ \ \ \ \ \ \ \ \ \ \ +\frac{1}{2} f X^\prime \phi^\prime g_{\alpha \beta} - X^\prime \delta^r_{( \alpha} \delta^r_{\beta ) } \phi^\prime \bigg] .  
\end{split}
\end{equation}
noting that $G_\alpha = \partial_\alpha G = G_X \partial_\alpha X$. The $tt$-component of the stress-energy tensor yields
\begin{equation}
\label{eq:stress_energy_scalar_tt}
\begin{split}
8 \pi T^{(\phi)}_{tt} =& -\frac{1}{2} \left( -\frac{1}{2} f \phi^{\prime 2} + L \right) h \\
&+ \frac{1}{2} f G_X h \phi^\prime \left(\frac{1}{2} f^\prime \phi^{\prime 2} + f \phi^\prime \phi^{\prime \prime} \right)  .  
\end{split}
\end{equation}
so that the $tt$-component of the Einstein equation becomes
\begin{equation}
\begin{split}
-\frac{h f^{\prime}}{r} -\frac{f h}{r^2}&+\frac{h}{r^2} + \frac{1}{2} \left( -\frac{1}{2} f \phi^{\prime 2} + L \right) h \\
& - \frac{1}{2} f G_X h \phi^\prime \left(\frac{1}{2} f^\prime \phi^{\prime 2} + f \phi^\prime \phi^{\prime \prime} \right) = 0 .
\end{split}
\end{equation}
Multiplying this result by $-4r^2 /h$, then we obtain Eq. (\ref{eq:einstein_tt}). The $rr$-component of the stress-energy tensor yields
\begin{equation}
\label{eq:stress_energy_scalar_rr}
8 \pi T^{(\phi)}_{rr} = \frac{1}{4} \phi^{\prime 2} + \frac{L}{2f} - \frac{1}{4} f G_X \phi^{\prime 3} \left( \frac{h^\prime}{h} + \frac{4}{r} \right) .
\end{equation}
so that the $rr$-component of the Einstein equation becomes
\begin{equation}
\begin{split}
&-\frac{1}{r^2 f}+\frac{h^{\prime}}{r h}+\frac{1}{r^2} \\
&- \frac{1}{4} \phi^{\prime 2} - \frac{L}{2f} + \frac{1}{4} f G_X \phi^{\prime 3} \left( \frac{h^\prime}{h} + \frac{4}{r} \right) = 0 .
\end{split}
\end{equation}
Multiplying the above expression by $4 f r^2$ leads to Eq. (\ref{eq:einstein_rr}).

\section*{Acknowledgements}
The authors are grateful to Shuang-Yong Zhou, Kristian Hauser Villegas, and Jan Tristram Acu\~{n}a for constructive criticism of the preliminary version of the manuscript. This research is supported by the University of the Philippines OVPAA through Grant No.~OVPAA-BPhD-2016-13. 

\bibliographystyle{apsrev4-1}
\bibliography{solutions_horndeski}

\end{document}